\documentclass[a4paper,unpublished,twocolumn,11pt]{quantumarticle}

\pdfoutput=1
\usepackage[utf8]{inputenc}
\usepackage[T1]{fontenc}
\usepackage{amsmath, amssymb, amsthm, physics}
\usepackage[countmax]{subfloat}

\usepackage{bbm}
\usepackage{graphicx}

\usepackage{hyperref}
\usepackage{epstopdf}
\usepackage{dsfont}
\usepackage{wrapfig}
\usepackage[caption=false]{subfig}
\usepackage{relsize}

\usepackage{tikz} 
\usetikzlibrary{calc, arrows, patterns,shapes, decorations, arrows.meta, fit, positioning}
\tikzset{auto,node distance =1 cm and 1 cm,semithick,
    var/.style ={circle, draw, minimum width = 1cm, ultra thick},
    latent/.style ={regular polygon, regular polygon sides=3, inner sep=1pt, draw, minimum width = 1.2cm, ultra thick},
    point/.style = {circle, draw, inner sep=0.06cm, fill, node contents={}},
    triangle/.style = {regular polygon, regular polygon sides=3, draw, inner sep=0.06cm, fill, node contents={}},
    bidir/.style={Latex-Latex,dashed},
    dir/.style={-Latex, thick},
    el/.style = {inner sep=2pt, align=left, sloped}}
\tikzstyle{vertex}=[circle, fill=black!10, draw=black]
\tikzstyle{edge}=[thick]
\tikzstyle{clique}=[line width=4, draw=black!70]

\graphicspath{{Images/}}
\DeclareMathOperator{\QSTAB}{QSTAB}
\DeclareMathOperator{\STAB}{STAB}
\DeclareMathOperator{\TBODY}{TH}

\DeclareMathOperator{\HSTAB}{HSTAB}
\DeclareMathOperator{\chull}{conv}
\DeclareMathOperator{\DAG}{DAG}
\DeclareMathOperator{\Ci}{Ci}
\DeclareMathOperator{\vertex}{vert}

\newtheorem{theo}{Thm.}
\newtheorem{prop}{Proposition}

\newcommand{\NS}{\mathrm{NS}}
\newcommand{\floor}[1]{\lfloor #1 \rfloor}
\newtheorem*{defi}{Def.}

\begin{document}

\title{Characterizing Hybrid Causal Structures with the Exclusivity Graph Approach
}
\author{Giovanni Rodari}
\affiliation{Dipartimento di Fisica, Sapienza Universit\`{a} di Roma, Piazzale Aldo Moro 5, I-00185 Roma, Italy}
\author{Davide Poderini}
\affiliation{Dipartimento di Fisica, Sapienza Universit\`{a} di Roma, Piazzale Aldo Moro 5, I-00185 Roma, Italy}
\affiliation{International Institute of Physics, Federal University of Rio Grande do Norte, 59078-970, P. O. Box 1613, Natal, Brazil}
\author{Emanuele Polino}
\affiliation{Dipartimento di Fisica, Sapienza Universit\`{a} di Roma, Piazzale Aldo Moro 5, I-00185 Roma, Italy}
\affiliation{Centre for Quantum Dynamics and Centre for Quantum Computation and Communication Technology, Griffith University, Yuggera Country, Brisbane, Queensland 4111, Australia}
\author{Alessia Suprano}
\affiliation{Dipartimento di Fisica, Sapienza Universit\`{a} di Roma, Piazzale Aldo Moro 5, I-00185 Roma, Italy}
\author{Fabio Sciarrino } 
\email{fabio.sciarrino@uniroma1.it}
\affiliation{Dipartimento di Fisica, Sapienza Universit\`{a} di Roma, Piazzale Aldo Moro 5, I-00185 Roma, Italy}
\author{Rafael Chaves}
\email{rafael.chaves@ufrn.br}
\affiliation{International Institute of Physics, Federal University of Rio Grande do Norte, 59078-970, P. O. Box 1613, Natal, Brazil}
\affiliation{School of Science and Technology, Federal University of Rio Grande do Norte, Natal, Brazil}

\begin{abstract}
Analyzing the geometry of correlation sets constrained by general causal structures is of paramount importance for foundational and quantum technology research. 
Addressing this task is generally challenging, prompting the development of diverse theoretical techniques for distinct scenarios. Recently, novel hybrid scenarios combining different causal assumptions within different parts of the causal structure have emerged. 
In this work, we extend a graph theoretical technique to explore classical, quantum, and no-signaling distributions in hybrid scenarios,
where classical causal constraints and weaker no-signaling ones are used for different nodes of the causal structure. 
By mapping such causal relationships into an undirected graph 
we are able to characterize the associated sets of compatible distributions and analyze their relationships.
In particular we show how with our method we can construct minimal Bell-like inequalities capable of simultaneously distinguishing classical, quantum, and no-signaling behaviors, and efficiently estimate the corresponding bounds.
The demonstrated method will represent a powerful tool to study quantum networks and for applications in quantum information tasks.
\end{abstract}

\maketitle

\section{Introduction}


\noindent Bell nonlocality~\cite{bell1964einstein,brunner2014bell,scarani2019bell} represents one of the most striking departures of quantum mechanics from our classical intuition.
Since its inception, much work has been devoted to the development of theoretical concepts and computational tools to study this phenomenon~\cite{brunner2014bell,scarani2019bell,wolfe2020quantifying,tavakoli2023SemidefiniteProgrammingRelaxations}, driven both by its foundational content as well as its applications for quantum information processing~\cite{gisin1991bell,gisin2007quantum,bera2017randomness,acin2016certified,vsupic2020self}. The violation of Bell's inequalities and more in general causal constraints, is a powerful tool to assess the nonclassicality of quantum systems in the so-called device-independent scenario where no assumptions are needed on the inner working of the employed apparatuses \cite{scarani20124,poderini2022ab}.
Notably, the concept of quantum nonlocality has been extended to more complex scenarios involving multiple parties \cite{bancal2013definitions,chaves2017causal} and general network configurations~\cite{branciard2012bilocal, tavakoli2014nonlocal, tavakoli2016bell,chaves2016entropic, chaves2016polynomial,rosset2016nonlinear, chaves2018quantum,tavakoli2021bell,chaves2021causal,polino2023experimental}.
This was possible also thanks to the realization that Bell's theorem \cite{bell1964einstein} can be seen as a particular instance of the incompatibility between quantum mechanics and classical causal reasoning~\cite{wood2015lesson,chaves2015unifying,pearl2009causality,wiseman2017causarum}.

Recently~\cite{bowles2020single, boghiu2021device, temistocles2019measurement,mazzari2023generalized,villegasaguilar2023nonlocality}, particular causal scenarios featuring additional compatibility constraints on the parties have been considered. 
The so-called \textit{broadcasting scenario} can be thought of as an extension of the usual Bell bipartite scenario, where one of the two parties is split into two separated sub-parties on which
one further imposes only weaker no-signaling constraints instead of the ones given by the assumption of local causality.
In this context, it is essential to devise methods capable of characterizing general scenarios with their hybrid constraints. Specifically, a significant inquiry revolves around giving a description of the properties of probability distributions compatible with such scenarios, together with determining the instances where the limitations imposed by hybrid constraints resemble what is found for multipartite Bell scenarios.

In this work, we show how the exclusivity graph (EG) approach, originally developed for the study of noncontextuality and nonlocality scenarios~\cite{cabello2010non,cabello2014graph,cabello2013basic,cabello2013simple,sadiq2013bell,rabelo2014multigraph,bharti2020uniqueness,poderini2020exclusivity,acin2015combinatorial,cabello2019quantum,bharti2021graph} and later applied to a larger class of causal structures~\cite{poderini2020exclusivity}, can be further extended to study hybrid scenarios like the \emph{broadcasting scenario} and its generalizations.
The main purpose of the EG method is to efficiently obtain constraints on the probabilistic distribution associated with the given causal scenario, both for classical (local) theory and quantum theory. 
To this goal one associates to a given causal scenario an undirected graph, called \emph{exclusivity graph}, in which vertices represent \emph{events} and edges represent \emph{exclusivity relations} between them, thus bridging elements from probabilistic theory into elements of graph theory.
To analyze a \emph{generalized} broadcasting scenario, we will extend the EG framework by introducing a construction that will allow us to embed the relaxed causal assumptions made on one part of the model into \emph{hybrid}-defined graph sets. This mapping will enable us to recover a description of the properties of probability distributions associated with such scenarios given their hybrid constraints.
Thus, this extension provides a useful tool to analyze such hybrid scenarios, giving a picture of the main features associated with the probabilistic behaviors arising in such structures; and providing a method to construct Bell-like inequalities in a graph-theoretical way, also recovering their associated numerical bounds.

The paper is organized as follows: in Sec.\ref{sec:broad} the main features related to the broadcasting scenario from the point of view of nonlocality theory will be presented, followed in Sec.\ref{sec:egapp} by a description of the basic ideas about the exclusivity graph approach and its connection with the description of causal scenarios. Then in Sec.\ref{sec:egextend} we will describe how such an approach can be extended and generalized to be applied to hybrid scenarios, and how one can infer properties of the hybrid constraints and obtain related Bell-like inequalities. 
Lastly, in Sec.\ref{sec:examples}, we construct two significant instances of minimal Bell-like inequalities with quantum violation demonstrating the no trivial relationships between the sets of compatible distributions in a broadcasting scenario.

\section{The broadcasting scenario}
\label{sec:broad}

\begin{figure}[t]
\centering
\subfloat[Bipartite Bell scenario.]{
    \centering
    \begin{tikzpicture}
         \node[var] (a) at (-2,0) {$A$};
         \node[var] (x) [above =of a] {$X$};
         \node[latent] (l) at (0,0) {$\Lambda$};
         \node[var] (b) at (2,0) {$B$};
         \node[var] (y) [above =of b] {$Y$};

         \path[dir] (x) edge (a) (y) edge (b); 
         \path[dir] (l) edge (a) (l) edge (b);
    \end{tikzpicture}
    \label{fig:bell}}
    \quad
\subfloat[Tripartite Broadcasting scenario.]{
    \centering
    \begin{tikzpicture}
         \node[var] (a) at (-2,0) {$A$};
         \node[var] (x) [above =of a] {$X$};
         \node[latent] (l) at (0,0) {$\Lambda$};
         \node[latent] (s) at (2,0) {$\Sigma$};
         \node[var] (b1) at (3,1) {$B_1$};
         \node[var] (b2) at (3,-1) {$B_2$};
         \node[var] (y1) [right =of b1] {$Y_1$};
         \node[var] (y2) [right =of b2] {$Y_2$};

         \path[dir] (x) edge (a) (y1) edge (b1) (y2) edge (b2);
         \path[dir] (l) edge (a) (l) edge (s);
         \path[dir] (s) edge (b1) (s) edge (b2);
         \node[thick, draw=black!70, color=black!70, dashed, fit=(s) (y1) (y2), inner sep=0.4cm] (ns)
         {$\NS$};
    \end{tikzpicture}
    \label{fig:broadcast}}
\caption{\textbf{The Broadcasting causal scenario:} \textbf{(a)} In the usual bipartite Bell scenario a classical source $\Lambda$ is shared between two separated parties, Alice and Bob, whose measurements settings and outcomes are represented respectively by pairs of discrete random variables $(X, A)$ and $(Y, B)$. \textbf{(b)} In the Broadcasting scenario part of the classical system is \textit{broadcasted} through a device $\Sigma$, to two distinct parties which perform local measurements on the subsystem received. $\Lambda$ still represents a source of classical correlations, while $\Sigma$ is assumed to be a general no-signaling latent variable causally dependent on $\Lambda$.}
\label{fig:bell_n_broadcast}
\end{figure}
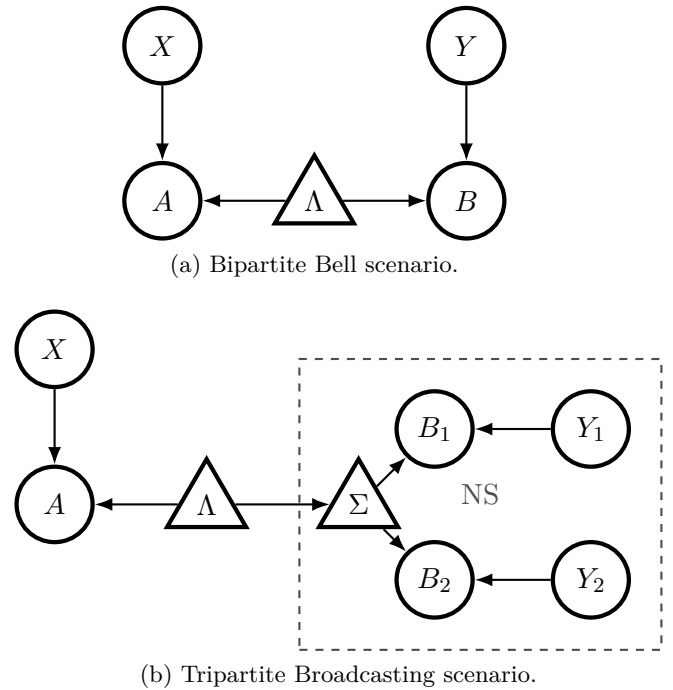

In the usual bipartite Bell scenario, whose causal model is depicted in Fig.\ref{fig:bell}, the assumption of local causality corresponds to the fact that any \textit{local} probability distribution compatible with such scenario must admit a decomposition in the form~\cite{brunner2014bell}
\begin{equation}
    p_L(ab|xy) = \sum_{\lambda} p(\lambda)p(a|x,\lambda)p(b|y,\lambda) \, \in \mathcal{L},
    \label{eqn:bell_local}
\end{equation}
where $\mathcal{L}$ represents a convex polytope~\cite{goh2018geometry}, in probability space, containing all the local distributions.
Conversely, quantum mechanics allows for distributions in the form
\begin{equation}
     p_Q(ab|xy) = \tr(\rho \Pi_{a|x} \otimes \Pi_{b|y}) \in \mathcal{Q},
     \label{eqn:bell_quantum}
\end{equation}
which belong to a convex set\cite{goh2018geometry}, namely $\mathcal{Q}$, that, as shown by Bell's theorem \cite{bell1964einstein}, is strictly larger than $\mathcal{L}$, that is, $\mathcal{L} \subset \mathcal{Q}$.
The particular class of quantum resources $\rho$ for which there exist measurements described by POVM elements $\{\Pi_{a|x}, \Pi_{b|y}\}$ generating $p_Q \notin \mathcal{L}$ are called Bell-nonlocal states \cite{brunner2014bell}. Notably, while all \textit{pure} entangled bipartite states can lead to Bell-nonlocal behaviors \cite{gisin1991bell}, the same cannot be said for \textit{mixed} entangled states, as first noted by Werner \cite{werner1989quantum}. This fact becomes a limitation in practical realizations of quantum information protocols where a direct measurement of a nonlocal behavior, by means of a witness such as the violation of the CHSH inequality \cite{clauser1969proposed}, is used to \textit{certify} for the presence of a quantum resource. 

In a recently proposed scenario~\cite{bowles2020single, boghiu2021device}, whose causal structure is depicted in Fig.~\ref{fig:broadcast}, one introduces a device $\Sigma$ \footnote{which in practical terms could be represented by a quantum channel where Bob's share of the system interacts with an ancilla state.}, which \textit{broadcasts} Bob's subsystem to two parties, here $B_1$ and  $B_2$.
In this case, one introduces the concept of a \textit{broadcast-local} behaviour as a probability distribution $p(ab_1 b_2|x y_1 y_2)$ which can be written as
\begin{multline}
    p(ab_1b_2|xy_1y_2) = \\
    = \sum_\lambda p(\lambda) p(a|x,\lambda) p^{NS}_B(b_1b_2|y_1y_2,\sigma_\lambda).
    \label{eqn:broadcast}
\end{multline}
Note that this embeds a \textit{local causality} condition between the partitions $\{A\}, \{B_1, B_2\}$, while between $\{B_1\}$ and $\{B_2\}$ one allows $p^{NS}_{B}(b_1 b_2|y_1 y_2,\sigma_\lambda)$ to be an arbitrary non-signaling distribution, i.e. a distribution constrained only by
\begin{equation}
\begin{split}
    \sum_{b_1} p^{NS}_{B}(b_1 b_2|y_1 y_2,\sigma_\lambda) = \sum_{b_1} p^{NS}_{B}(b_1 b_2|y_1' y_2,\sigma_\lambda) \quad \\ \forall \; b_2,y_2,y_1,y_1',\sigma_\lambda, \\
    \sum_{b_2} p^{NS}_{B}(b_1 b_2|y_1 y_2,\sigma_\lambda) = \sum_{b_2} p^{NS}_{B}(b_1 b_2|y_1 y_2',\sigma_\lambda) \quad \\ \forall \; b_1,y_1,y_2,y_2',\sigma_\lambda.
    \label{eqn:NS1}
\end{split}
\end{equation}
Equation \eqref{eqn:broadcast} defines a set of distributions that is strictly larger than the \textit{local} set of distributions in the tripartite Bell scenario~\cite{sliwa2003symmetries}.
Such behaviors are relevant since by imposing the weaker no-signaling constraint on $p^{NS}_B(b_1 b_2|y_1 y_2,\sigma_\lambda)$, a quantum violation of \eqref{eqn:broadcast}, through distributions in the form
\begin{multline}
    p_Q(a b_1 b_2|x y_1 y_2) = \\
    = \tr(\rho_{AB_1 B_2} \Pi_{a|x} \otimes \Pi_{b_1|y_1} \otimes \Pi_{b_1|y_1}),
\end{multline}
on states obtained as
\begin{equation}
    \rho_{AB_1 B_2} = (\mathds{1}_A \otimes \Sigma_{B \rightarrow B_1 B_2}) \rho_{AB},
\end{equation}
cannot originate from the transformation device alone. This can be used to prove the nonclassicality of bipartite mixed states which are Bell-local, thus leading to a previously unknown form of single-copy activation~\cite{bowles2020single, boghiu2021device}, as recently experimentally demonstrated in \cite{villegasaguilar2023nonlocality}. 

\section{Exclusivity graph approach}
\label{sec:egapp}
In what follows, we briefly describe the \textit{exclusivity graph} (EG) method, a graph-theoretical approach that has been applied to the study both of noncontextual and nonlocal scenarios to obtain useful properties of the related local and quantum behaviors sets~\cite{cabello2010non,cabello2014graph,cabello2013basic,cabello2013simple,sadiq2013bell,rabelo2014multigraph,bharti2020uniqueness,poderini2020exclusivity,acin2015combinatorial,cabello2019quantum,bharti2021graph}.

For simplicity, one can start from the causal model depicted in Fig.~\ref{fig:bell}, which represents a standard bipartite Bell scenario $(n,m;k,j)$, where $(n,m)$ and $(k,j)$ indicate respectively the number of possible settings and outcomes for Alice and Bob. 
Within the EG approach, an exclusivity graph is an undirected graph $G^{(n,m,k,j)} = (V,E)$ in which each vertex $v \in V$ represents an event $v \equiv (ab|xy)$, i.e. a pair of measurement outcomes $\{a,b\}$ given a pair of measurement settings $\{x,y\}$. Two vertices $u,v \in V$ are connected by an edge, i.e. they are adjacent, if and only if they are mutually exclusive. Intuitively, by exclusive events, we mean that they cannot be simultaneously observed in the same run of an experiment.
Formally, two events $u \equiv (ab|xy)$ and $v \equiv (a'b'|x'y')$, are exclusive, that is $(u,v) \in E$, if and only if at least one of the following two conditions holds:
\begin{gather*}
    x = x' \; \textrm{and} \; a \neq a', \\
    y = y' \; \textrm{and} \; b \neq b'.
\end{gather*}

Bell inequalities, implied by the local causality assumption \eqref{eqn:bell_local}, can generically be expressed as
\begin{equation}
    I_w(p) = \sum_{\substack{a,b \\ x,y}} w_{ab|xy} \; p(ab|xy) \leq \beta_L.
\end{equation}
In the EG approach, the expression $I_w(p)$ can be mapped into a graph by associating a set of real numbers $\{w_{v}\}$, called weights, to the corresponding vertices in $G$, so that
\begin{equation}
    I_w(p) \equiv I(G,w) = \sum_{v \in G} w_v p(v).
    \label{eqn:gr_ineq}
\end{equation}
Given an undirected graph $G = (V,E)$, we call a \emph{labeling} an assignment $L:V\rightarrow \mathbb{R}^d$ of a real number, or a vector, to each node in $G$.
In particular we can define the following labelings~\cite{knuth1993sandwich}:
\begin{itemize}
    \item a \textit{characteristic labeling} of a subset $U \subseteq V(G)$, which is a vector $\Vec{x}$ indexed by vertices $v \in V$ such that $x_v = 1$ if $v \in U$ and $x_v = 0$ otherwise.
    \item an \textit{orthonormal labeling}, which is a map $a_v: V \rightarrow \mathbb{R}^d$ such that $a_v \cdot a_u = 0$ if $(u,v) \in E(G)$ and $|a_v|^2 = 1$.
    \item a \textit{positive labeling}, which is a map  $a_v: V \rightarrow \mathbb{R}^+$ which associates a positive real number to each vertex $v \in V(G)$.
\end{itemize}
\begin{widetext}
From them, we can construct the following sets associated to a graph $G$3
\begin{align*}
    \text{STAB}(G) &= \chull\{\Vec{x} : \Vec{x} \; \text{is a characteristic labeling of a stable set of} \; G\}, \\
    \text{TH}(G) &= \{\Vec{x} : x_v = (a_v)_1 \; a_v \; \text{is an orthonormal labeling of} \; G\}, \\
    \text{QSTAB}(G) &= \{\Vec{x} \; \text{is a positive labeling} : \sum_{v \in Q} x_v \leq 1 \; \text{for all cliques Q of G}\},
\end{align*}
where a \textit{stable set} is a subset $S \subseteq V(G)$ such that for each pair of vertices $u,v \in S$ then $(u,v) \notin E(G)$; and a \textit{clique} is a subset $Q \subseteq V(G)$ containing all mutually adjacent vertices. Notably, both $\STAB(G)$ and $\QSTAB(G)$ are convex polytopes, since the number of both stable sets and cliques in a graph $G$ are finite; while $\mathrm{TH}(G)$ is just a convex set.
\end{widetext}
Given a set of real weights $\{w_{v}\}$ for $G$ one can then define the following \textit{graph invariant} quantities~\cite{knuth1993sandwich}:
\begin{align}
    \label{eq:alpha_w}
    \alpha(G,w) &= \max\{w \cdot x : x \in \text{STAB}(G)\},\\
    \label{eq:theta_w}
    \theta(G,w) &= \max\{w \cdot x : x \in \text{TH}(G)\},\\
    \label{eq:alpha_starw}
    \alpha^*(G,w) &= \max\{w \cdot x : x \in \text{QSTAB}(G)\},
\end{align}
Notably $\alpha(G,w)$ and $\theta(G,w)$, known as \textit{independence number} and \textit{Lova\`sz Theta} of a graph~\cite{knuth1993sandwich,grotschel1986relaxations}, give respectively a tight local bound (LHV) and an upper quantum bound (Q) for the inequality $I_w(p)$ when it is mapped onto an exclusivity graph $G$ and a set of weights $w$ ~\cite{cabello2013simple, cabello2013basic, poderini2020exclusivity}. 
That is, one can write:
\begin{equation}
    I_w(p) \equiv I(G,w) \overset{\text{LHV}}{\leq} \alpha(G,w) \overset{\text{Q}}{\leq} \theta(G,w) .
\end{equation}
That this is true for the LHV bound can be seen by noticing that $\STAB(G)$ constitutes a convex polytope whose extremal points are exactly all the possible deterministic assignments respecting the exclusivity constraints embedded in the edges of the graph $G$. 
The validity of the quantum bound instead, can be intutively explained recalling that exclusivity relations between vertices in $G$ translate, in a quantum mechanical framework, into \textit{orthogonality} between projectors~\cite{fritz2013local,sainz2014exploring,amaral2018graph,poderini2020exclusivity}. Thus by construction $\text{TH}(G)$ includes the set of all quantum correlations compatible with the scenario under consideration.
This inclusion is not strict: indeed it can be shown that it includes a slightly larger set, called \textit{almost quantum correlations}~\cite{navascues2015almost}. Nonetheless, $\theta(G,w)$ reproduces exactly the quantum bound for some known Bell inequalities such as the CHSH inequality~\cite{cabello2013basic,sadiq2013bell,bharti2021graph}.

The last graph invariant left to consider is $\alpha^*(G,w)$, connected with $\QSTAB(G)$: in general, the constraints defining $\QSTAB$ are related to the notion known as \emph{exclusivity principle} (EP)~\cite{cabello2013simple, cabello2013basic, poderini2020exclusivity} or \emph{local orthogonality}. Due to the definition of $\QSTAB$, at its core, the exclusivity principle 
defines a constraint on the admitted probability distributions which is generally weaker than the one given by the orthogonality relations of quantum mechanics. That is, with the EP one can describe a correlation set that is larger than the quantum set, even when dealing with the mapping of a linear inequality onto a weighted graph $(G,w)$.
Notably, when one describes a Bell bipartite scenario in the EG framework, the set described by $\QSTAB$ corresponds to the \emph{no-signaling} behavior set in nonlocality theory~\cite{cabello2014graph,acin2015combinatorial}, meaning that $\alpha^*(G,w)$ represents the no-signaling bound for an inequality $I(G,w)$.

Crucially, the previous reasoning can be applied to any unitary weighted ($w_v = 1 \; \forall v$) \emph{induced} subgraph $H$ of $G$, i.e. a graph $H$ such that $V(H) \subseteq V(G)$ and $(u,v) \in E(H) \Leftrightarrow (u,v) \in E(G)$. This is equivalent to considering a subset of the original graph and a simpler linear inequality whose constraints, embedded in its edges, are inherited from the original exclusivity graph but for which the relevant graph invariants are typically easier to compute:
\begin{equation}
    I(H) = \sum_{v \in H}p(v) \overset{\text{LHV}}{\leq} \alpha(H) \overset{\text{Q}}{\leq} \theta(H) \overset{\text{EP(NS)}}{\leq} \alpha^*(H).
    \label{eqn:IH}
\end{equation}

\section{Exclusivity Graphs in Hybrid Causal Scenarios}
\label{sec:egextend}
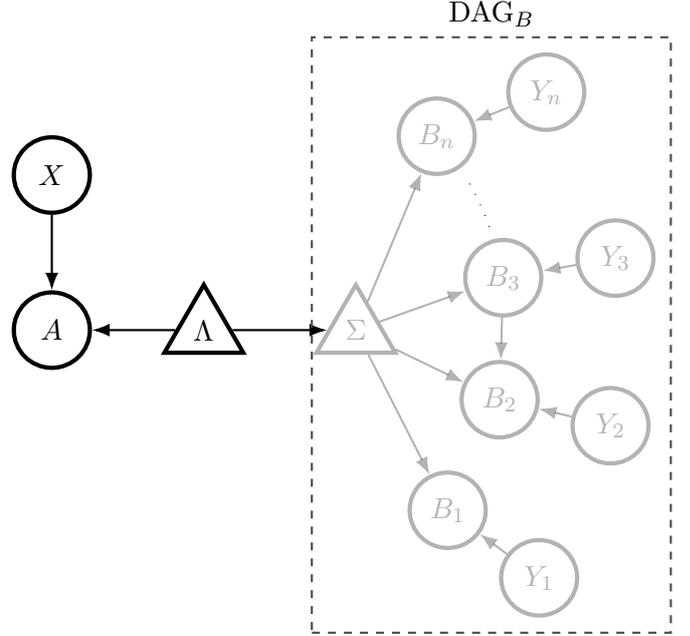
\begin{figure}[h]
    \centering
    \begin{tikzpicture}
         \node[var] (a) at (-2,0) {$A$};
         \node[var] (x) [above =of a] {$X$};
         \node[latent] (l) at (0,0) {$\Lambda$};

         \node[latent,  opacity=.3] (s) at (2,0) {$\Sigma$};
        \foreach \i in {1, 2, 3}{
            \draw (\i*70/3-60: 4cm) node[var, opacity=.3] (b\i) {$B_\i$};
            \draw (\i*70/3-60: 5.5cm) node[var, opacity=.3] (y\i) {$Y_\i$};
			\path[dir,  opacity=.3] (s) edge (b\i) (y\i) edge (b\i);
        }
         \draw (40: 4cm) node[var, opacity=.3] (bn) {$B_n$};
         \draw (35: 5.5cm) node[var, opacity=.3] (yn) {$Y_n$};
        \draw [loosely dotted] (20:4)  arc[radius = 4cm, start angle= 20, end angle = 30];
         \path[dir, opacity=.3] (s) edge (bn) (yn) edge (bn) (b3) edge (b2);

         \path[dir] (x) edge (a);
         \path[dir] (l) edge (a) (l) edge (s);
         \node[thick, draw=black!70, color=black!70, dashed, fit=(s) (y1) (y3) (yn), inner sep=0.2cm, label=$\mathrm{DAG}_B$] (dagb) {};
    \end{tikzpicture}
     \caption{\textbf{Hybrid causal structure:} starting from a bipartite Bell scenario, on Bob's side the latent node $\Sigma$ broadcasts the system produced by $\Lambda$ to a fixed causal structure, denoted as $\DAG_B$, described by two sets of variables $\{B_i\}$ and $\{Y_i\}$. The goal is to derive constraints given by a \textit{global} local causality assumption between $A$ and $\mathrm{DAG}_B$, while only accounting for weaker \textit{no-signaling} constraints, or their generalization as the \textit{exclusivity principle}, between the variables contained in $\mathrm{DAG}_B$.}%
    \label{fig:dagb}
\end{figure}

Let us consider the model depicted in Fig. \ref{fig:dagb}, which we present as a generalization of the broadcasting scenario studied in~\cite{bowles2020single,boghiu2021device}. We wish to show how in the EG framework one can devise a construction able to capture the constraints of such a scenario using the local causality assumption on the global scenario between $A$ and $B$, together with the no-signaling constraint - or the generalization thereof in the exclusivity principle assumption - on the causal structure described by $\DAG_B$.

Similarly to equation~\eqref{eqn:broadcast}, the \textit{broadcasting-local} behaviour set which we associate to the general model in Fig.~\ref{fig:dagb} and wish to analyze is given by
\begin{multline}
    p(a,b_1,\ldots,b_n|x,y_1,\ldots,y_n) =
    \sum_{\lambda} p(\lambda) p(a|x,\lambda) \cdot\\
    \cdot p^{\mathrm{EP}}_{\DAG_B}(b_1,\ldots,b_n|y_1,\ldots,y_n,\sigma_\lambda) \, ,
    \label{eqn:broadcast_general}
\end{multline}
where $p^{\mathrm{EP}}_{\DAG_B}(\cdot)$ represents now a distribution fulfilling the EP. While the idea of such multi-node structures is present in~\cite{boghiu2021device}, we here admit, in principle, the presence of directed edges between the observed nodes of $\DAG_B$. In such case, due to the presence of direct a causal dependence between parties, the notion of no-signaling needs to be extended. We chose, as its generalization, the exclusivity principle, noticing that this constraint will naturally reduce to no-signaling when there are no direct causal edges, as in the causal scenario depicted in Fig.~\ref{fig:broadcast}.

\begin{figure*}[ht!]
    \centering
    \includegraphics[width=0.95\textwidth]{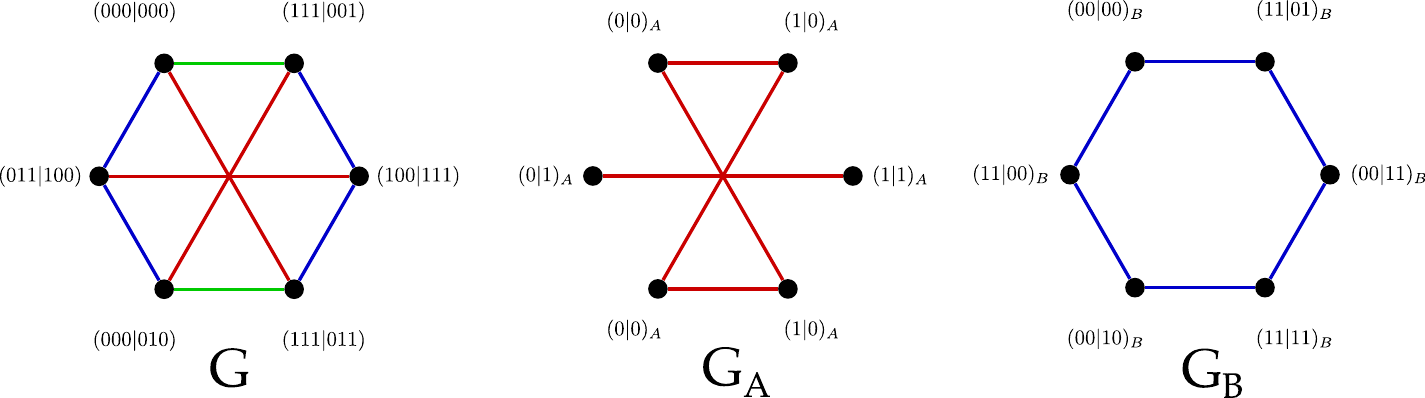}
    \caption{\textbf{Constructing $G_A$ and $G_B$ from an exclusivity graph $G$:} on the left we depict the exclusivity graph $G$ associated to a subset of events of the model in Fig. \ref{fig:broadcast}, where we assume $\{A,X,B_1,B_2,Y_1,Y_2\} \in \{0,1\}$. Each node is associated to an event $ab_1b_2|xy_1y_2$, and the edges are color-coded: the red edges connect exclusive events due to $a|x \sim a'|x'$, the blue edges due to $b_1b_2|y_1y_2 \sim b'_1b'_2|y'_1y'_2$, while the green edges are due to both. One can thus define $G_A$, depicted in the upper right, as the graph defined on the same vertices inheriting the edges depicted in red and green, related to $\{A,X\}$. Analogously, $G_B$ is constructed by accounting for the edges depicted in blue and green in $G$, related to $\DAG_B$.}
    \label{fig:bell_mapping}
\end{figure*}

The main idea behind our approach is to consider in a hybrid fashion the exclusivity relations on the two sides of the scenario. Extending the EG framework, given the general causal model as in Fig.~\ref{fig:dagb} one can construct the corresponding exclusivity graph $G = (V_G,E_G)$, and notice that, if two vertices $u,v \in V_G$ are adjacent in $G$, it could be due to an exclusivity relation given by the parameters describing Alice's side of the causal scenario (e.g $a|x$ and $a'|x$ with $a \neq a'$ w.r.t. Fig.~\ref{fig:dagb}), to an exclusivity relation described by parameters belonging to $\DAG_B$, or both. Thus, we can split the edge set $E_G$ into two subsets $E_A$ and $E_B$ defined as
\begin{align}
    \nonumber
    E_A &= \{(u,v) \in E_G \;|\; u,v \; \text{exclusive w.r.t. } A,X \}, \\
    E_B &= \{(u,v) \in E_G \;|\; u,v \; \text{exclusive w.r.t.} \DAG_B \}.
\end{align}
In such a way, one defines two graphs $G_A = (V_G, E_A)$ and $G_B = (V_G, E_B)$, which in general are not disjoint ($E_A \cap E_B \neq \emptyset$), connected to the exclusivity relations inferred respectively on $A,X$ and $\DAG_B$, such that
\begin{equation}
\begin{split}
    G = (V_G,E_G); \;G_A = (V_A,E_A); \;  G_B = (V_B,E_B) \\
    \textrm{s.t.} \; V_G = V_A = V_B; \; E_G = E_A \cup E_B,
    \label{eqn:egext_def}
\end{split}
\end{equation}
as pictorially shown in Fig.~\ref{fig:bell_mapping}. Our goal is now to show how $G_A$, $G_B$, and the related graph sets can be used to derive information on the \textit{broadcasting-local} distributions as in Eqn.~\eqref{eqn:broadcast_general}. Again, we stress that we are considering a hybrid causal model, where the causal constraints imposed on \textit{some} subset of its nodes - e.g. $\DAG_B$ of Fig. \ref{fig:broadcast} - have been relaxed to the more general exclusivity principle, a weaker constraint than the one imposed by the algebra of quantum mechanics.

To derive properties of the \textit{broadcasting-local} distributions in a graph theoretical way, we introduce the following operation amongst convex polytopes defined over $\mathds{R}^d_+$:
\begin{multline}
    R := P \odot Q = \chull\{x \circ y \;|\; \forall (x,y) \\
    \textrm{s.t.} \; x \in \vertex(P); \; y \in \vertex(Q) \}
    \label{eqn:odot_poly}
\end{multline}
where with $x \circ y$ we denote the element-wise product amongst vectors, also known as Hadamard product~\cite{horn2012matrix}, while $\chull(\cdot)$ and $\vertex(\cdot)$ are respectively the convex hull and the finite set of extremal points of the given set, as defined in appendix~\ref{apx:matdefi}. 
Given this definition, in the following, we present the results obtained using the EG approach applied to hybrid causal scenarios.

\subsection{Recovering the local polytope}

To showcase the ideas behind our hybrid construction of correlation sets in such scenarios, consider, without loss of generality, the causal structure depicted in Fig.~\ref{fig:broadcast}.
Here, if $\Sigma$ is interpreted as a \textit{local} latent variable, then the structure is equivalent to a standard tripartite Bell scenario~\cite{svetlichny1987distinguishing}. In this case, we can write the corresponding \textit{local} distributions as
\begin{multline}
    p_L(ab_1b_2|xy_1y_2) = \\
    = \sum_\lambda p(\lambda)p(a|x,\lambda)p(b_1|y_1,\lambda)p(b_2|y_2,\lambda)\;.
    \label{eqn:trilocal_distrib}
\end{multline}
Such a constraint defines a convex polytope whose extremal points, due to Fine's Theorem \cite{fine1982hidden}, are given by 
\begin{equation}
    p_L(ab_1b_2|xy_1y_2) = D(a|x)D(b_1b_2|y_1y_2),
    \label{eqn:det_local}
\end{equation}
where $D(\cdot)$ represents an arbitrary \textit{deterministic} distribution~\footnote{With deterministic, here, we denote a function $D(a|x) = \delta_{f(x)}^a$ where $f(\cdot)$ is some function from the settings $x$ to the outcomes $a$.}.

In the proposed extension of the EG framework, one can consider the two graphs $G_A = (V_G, E_A)$ and $G_B = (V_G, E_B)$ as given in Eqn. \eqref{eqn:egext_def}. In this case, given two vertices $u,v \in V_G$, i.e. events $u \equiv ab_1b_2|xy_1y_2$ and $v \equiv a'b'_1b'_2|x'y'_1y'_2$, then
\begin{equation} 
    (u,v) \in E_A \; \Leftrightarrow \; a \neq a' \; \wedge \; x = x',
\end{equation}
and
\begin{equation} 
    (u,v) \in E_B \; \Leftrightarrow \; b_i \neq b_i' \; \wedge \; y_i = y_i'; \quad i = 1,2.
\end{equation}
Note that $G_A$ and $G_B$ embed respectively exclusivity constraints related to the causal parameters associated with different partitions of the events $(ab_1b_2|xy_1y_2)$. Both $\STAB(G_A)$ and $\STAB(G_B)$ are well defined and their extremal points are associated with the deterministic strategies $D(a|x)$ and $D(b_1b_2|y_1y_2)$, respectively. Then, as proven in the Appendix \ref{apx:prop1}, it follows that
\begin{prop}
    \begin{equation}
        \STAB(G) = \STAB(G_A) \odot \STAB(G_B).
    \end{equation}
    \label{prop:stabequiv}
\end{prop}

Essentially, this shows that, considering separately $G_A$ and $G_B$, one can recover $\STAB(G)$ and thus a description of the local polytope as in Eqn.~\eqref{eqn:trilocal_distrib}. Note that this mapping holds since $\STAB(G)$ is defined as the convex hull of all the possible stable labelings of $G$, which effectively represent all the possible \textit{deterministic} assignments given by \eqref{eqn:det_local}, compatible with the causal scenario under consideration and the exclusivity constraints embedded in the edges of the graph $G$.
That is, due to Prop.~\ref{prop:stabequiv}, we verify that our hybrid approach recovers, as it should, the description of a Bell-multipartite scenario, which is generated in a hybrid scenario when $\Sigma$ represents a local classical latent variable.

\subsection{Describing the broadcast-local behaviors}

We now use the same approach to describe the hybrid constraints. Here, the only constraint imposed amongst the nodes on $DAG_B$ is given by the exclusivity principle, thus generating distributions as in Eqn.~\eqref{eqn:broadcast}. Within the extended exclusivity graph framework then:
\begin{prop}
Given that both $\STAB(G_A)$ and $\QSTAB(G_B)$ are convex polytopes~\cite{knuth1993sandwich}, the set defined as:
\begin{equation}
    \HSTAB(G) \equiv \STAB(G_A) \odot \QSTAB(G_B),
    \label{eqn:hstab}
\end{equation}
reproduces the broadcasting-local set in a hybrid scenario.
\end{prop}
To show why this holds, for simplicity, we consider again the scenario depicted in Fig.~\ref{fig:broadcast}. Since the no-signaling distributions $p_{B}^{\NS}(b_1 b_2|y_1 y_2,\sigma_\lambda)$ considered here belong to a convex polytope, then the broadcasting-local behaviors themselves form a convex polytope. The extremal points of such polytope can be written as~\cite{bowles2020single} 
\begin{equation}
    E_{BR}(ab_1b_2|xy_1y_2) = D(a|x)N(b_1b_2|y_1y_2),
    \label{eqn:extremal_broad}
\end{equation}
where $N(b_1b_2|y_1y_2)$ represents an extremal point of the $\NS$ polytope on Bob's side, described by the constraints of Eqn. \eqref{eqn:NS1} when marginalized over $\sigma_\lambda$. 
Note that in this case, the causal structure $\DAG_B$ on Bob's side is equivalent to a bipartite Bell scenario, where, instead of local causality, only the exclusivity principle applies. While in general within the EG framework, the $\QSTAB$ set embeds a weaker constraint given by the \textit{exclusivity principle}, it has been shown that in this case, it is equivalent to the $\NS$ polytope~\cite{acin2015combinatorial}. Given this, we can associate the extremal points of $\QSTAB(G_B)$ to the extremal $\NS$ strategies $N(b_1b_2|y_1y_2)$: then, by its definition, $\HSTAB(G)$ will represent the convex hull of behaviors as in Eqn. \eqref{eqn:extremal_broad}, thus fully reproducing the broadcasting-local set.

The same reasoning holds with respect to any hybrid scenario. In the same way, the behaviors given by
\begin{multline*}
    p(a,b_1,\ldots,b_n|x,y_1,\ldots,y_n) = \\
    = \sum_{\lambda} p(\lambda) p(a|x,\lambda) p^{\mathrm{EP}}_{\DAG_B}(b_1,\ldots,b_n|y_1,\ldots,y_n,\sigma_\lambda),
\end{multline*}
can be seen as the convex hull of strategies in the form
\begin{multline*}
    E(a,b_1,\ldots,b_n|x,y_1,\ldots,y_n) = \\ = D(a|x) \mathrm{EP}_\mathrm{\DAG_B
    }(b_1,\ldots,b_n|y_1,\ldots,y_n).
\end{multline*}
In our construction, such strategies are effectively represented by the extremal points of the set $\HSTAB(G)$ defined as in \eqref{eqn:hstab}. In essence, this object will reproduce - in a graph-theoretical fashion - the broadcasting-local set with respect to any scenario described by the structure depicted in Fig.\ref{fig:dagb}. In the following, we will proceed to show how this mapping can be used to give an intuitive picture of some properties of the broadcasting-local set, together with a description of its relationship with the local set akin to a multipartite Bell scenario.

\subsection{Properties of the hybrid behavior set}

As shown, the defined set $\HSTAB(G)$ effectively describes the probability distributions compatible with a hybrid scenario, given the constraints arising from the assumption of \textit{global} local causality between the two sides of the causal model but considering only a weaker constraint given by the \textit{exclusivity principle} between the causal variables on one side of the scenario.

\begin{figure}[h]
    \centering
    \includegraphics[]{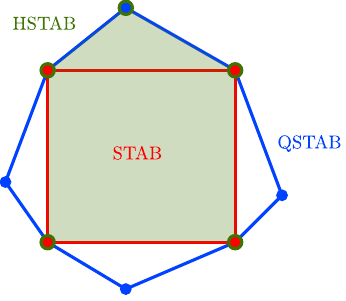}
    \caption{\textbf{Pictorial representation of $\HSTAB(G)$:} in this simplified two dimensional representation, $\STAB(G)$ corresponds to the square bounded by red lines, while $\QSTAB(G)$ to the polygon bounded by blue lines. The hybrid set $\HSTAB(G)$, associated with broadcast-local behaviors \eqref{eqn:broadcast_general}, is represented by the green area. Due to Prop.~\ref{prop:sandhstab}, $\STAB(G) \subseteq \HSTAB(G)$: since every extremal point of $\STAB(G)$ is still an extremal point of $\QSTAB(G)$, it follows that \textit{every} extremal point of $\STAB(G)$ is extremal for $\HSTAB(G)$. Also, $\HSTAB(G) \subseteq \QSTAB(G)$, and due to Prop. \ref{prop:hstabvert} in general not all extremal points of $\QSTAB(G)$ are extremal points for $\HSTAB(G)$.}
    \label{fig:hstab}
\end{figure}

We now wish to highlight the relationship between the broadcasting-local behaviors, represented in the EG framework by $\HSTAB(G)$, with the other behavior sets represented by $\STAB(G)$ (related to classical local distributions) and $\QSTAB(G)$ (related to distributions constrained by the exclusivity principle).
\begin{prop}
The extremal points of $\HSTAB(G)$ form a subset of the extremal points of $\QSTAB(G)$, that is,
\begin{equation*}
    \vertex(\HSTAB(G)) \subseteq \vertex(\QSTAB(G)).
\end{equation*}
\label{prop:hstabvert}
\end{prop}
This proposition (see appendix \ref{apx:prop2} for the proof) directly implies that the broadcast-local behaviors belong to the convex hull of deterministic strategies in the form
\begin{multline}
    D(ab_1\ldots b_n|x y_1\ldots y_n) = \\
    = \delta(a,f(x)) \prod_{i=1}^n \delta(b_i,g_i(\text{pa}(b_i))),
    \label{eqn:full_local}
\end{multline}
and \textit{some} of the extremal behaviors fulfilling the exclusivity principle
\begin{equation}
    \text{EP}(ab_1\ldots b_n|x y_1\ldots y_n) \in \vertex(\QSTAB(G)),
\end{equation}
both with respect to the \textit{whole} causal structure. 
Moreover, for a graph $G$ it holds $\STAB(G) \subseteq \QSTAB(G)$, a result known in graph theory as \textit{sandwich theorem}. This fact directly leads to the following result
\begin{prop}
The set $\HSTAB(G)$ is \textit{sandwiched} between $\STAB(G)$ and $\QSTAB(G)$, that is:
    \begin{equation*}
        \STAB(G) \subseteq \HSTAB(G) \subseteq \QSTAB(G).
    \end{equation*}
\label{prop:sandhstab}
\end{prop}
See appendix \ref{apx:prop3} for the proof.
These two results, combined together, show in a graph-theoretical way that due to the interpretation we have given to the graph sets of the exclusivity graph associated with a hybrid scenario, one has that the convex polytope of the broadcast-local behaviors as in Eqn.~\eqref{eqn:broadcast_general} defined by relaxing the causal constraints on $\DAG_B$:
\begin{enumerate}
     \item \textit{Strictly} contains the ones respecting a full local decomposition, obtained by considering $\Sigma$ as a local hidden variable, which can be expressed as a convex combination of \textit{deterministic} behaviors as in Eqn.~\eqref{eqn:full_local}.
     \item It is a subset of the behaviors fulfilling the \textit{exclusivity principle}, which can be seen as a generalization of the no-signaling constraint, with respect to the whole causal structure.
\end{enumerate}
This, in turn, gives a characterization of the geometrical properties of the broadcasting-local set in the behavior space, as pictorially summarized in Fig. \ref{fig:hstab}. Note that the result in Props. \ref{prop:hstabvert} and \ref{prop:sandhstab} can also be retrieved by considering the assumptions on the distributions that characterize these graph-theoretical sets: the exclusivity principle (ruling $\QSTAB(G)$) is strictly weaker than locality assumption (ruling $\STAB(G)$), while the broadcast-local set is encompassed by $\HSTAB(G)$ which is partially ruled by both.

\subsection{Bounding inequalities in hybrid scenarios}
\begin{figure}[ht]
    \centering
    \includegraphics[width=.8\columnwidth]{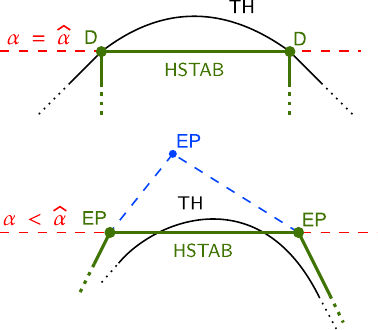}
    \caption{\textbf{Pictorial representation of inequalities in a hybrid scenario:} we give a two-dimensional representation of the possible inequalities in the hybrid scenario. At the top, we have depicted the case in which a facet of the $\HSTAB$, connecting two deterministic points, corresponds to a facet of the $\STAB$: this leads to an inequality for which $\alpha = \hat{\alpha} < \theta$, i.e. the local bound and the broadcasting-local bound coincide. On the bottom, an example of a genuine inequality that could detect the presence of a gap between the local set and the broadcasting-local set is depicted. A facet of $\HSTAB$ connects two extremal points fulfilling the exclusivity principle, cutting the quantum behavior set given by $\mathrm{TH}$: in this case, we would have that $\alpha < \hat{\alpha} < \theta$.}
    \label{fig:genuine}
\end{figure}

The construction we described so far can be applied to every induced subgraph $H$ of a full exclusivity graph $G$ associated with a hybrid scenario. In such a way, one can explore in a graph theoretical way the numerical bounds on linear probabilistic inequalities $I(H,w)$, mapped into a graph as in Eqn.~\eqref{eqn:IH}, given by different requirements at the level of allowed probability distributions. Hence, in this section, we will show how starting from the properties of the hybrid graph set defined previously, it is possible to derive conditions - based on the topology of the underlying graph - that have to be satisfied to demonstrate the equivalence of the broadcasting-local set with the local one and to possibly observe a quantum violation, i.e. a discrepancy between the broadcasting-local set and the quantum one.

To this purpose, consider an induced subgraph $H=(V_H, E_A, E_B)$ of $G$ which, following the prescriptions given before, can be decomposed into the following subgraphs:
\begin{equation}
    S_A = (V_H,E_A),\; S_B=(V_H,E_B), \; E_A \cup E_B = E_H,
\end{equation}
Similarly to Sec.\ref{sec:egapp}, to the hybrid graph set $\HSTAB(H)$ we will associate the following quantity:
\begin{equation}
     \hat{\alpha}(H,w) = \max\{w \cdot x : x \in \text{HSTAB}(H)\}.
\end{equation}
Using this quantity, one can then bound the linear inequality represented by the induced subgraph $H$ with respect to the broadcasting-local behaviors (B):
\begin{multline}
    I(H,w) = \sum_{v \in H}w_v p(v) \overset{\text{LHV}}{\leq} \alpha(H,w) \overset{\text{B}}{\leq} \\
    \overset{\text{B}}{\leq}  \hat{\alpha}(H,w) \overset{\text{EP(NS)}}{\leq} \alpha^*(H,w).
    \label{eqn:IB}
\end{multline}
where the chain of inequalities holds due to Prop.~\ref{prop:sandhstab}. 
We note that a quantum violation of $I(H,w)$ of the broadcast-local bound is possible if and only if
\begin{equation}
     \alpha(H,w) \leq \hat{\alpha}(H,w) < \theta(H,w).
\end{equation}
Let us recall a result known in graph theory as \textit{Strong Perfect Graph Theorem} \cite{lovasz1972characterization,chudnovsky2006strong}, which states
\begin{theo}[SPG]
A graph $G$ is \textit{perfect} if and only if it does not contain induced odd cycles or anti-cycles of cardinality $n \geq 5$. If a graph is perfect:
\begin{equation*}
    \STAB(G) = \TBODY(G) = \QSTAB(G),
\end{equation*}
and
\begin{equation*}
    \alpha(G,w) = \theta(G,w) = \alpha^*(G,w)\;.
\end{equation*}
\label{th:spgt}
\end{theo}
This leads to a necessary condition for $I(H,w)$~\footnote{The condition is not sufficient since in general the Lov\'asz Theta $\theta(H,w)$ also accounts for the so-called \textit{almost quantum correlations}, a set equivalent to the $1+AB$ level of the NPA hierarchy \cite{navascues2015almost}} to represent a linear inequality which could admit a quantum violation of a broadcasting-local bound: the {\em non-perfectness} of the associated graph $H$. This gives a simple criterion to identify whether an inequality could admit a quantum violation or not. Analogously, considering the topology of $S_A$ and $S_B$ we can identify some trivial limiting cases:
\begin{enumerate}
    \item If $E_A = \emptyset$, then $E_H = E_B$ and $\HSTAB(H) = \QSTAB(H)$, since $\STAB(S_A)$ contains the vector $\chi = (1,1,...,1)$. In this case, it follows that:
    \begin{equation}
         \hat{\alpha}(H,w) = \alpha^*(H,w),
    \end{equation}
    thus making $I(H,w)$ impossible to be violated by quantum correlations in the hybrid scenario.
    \item If $S_B$ is perfect, then $\STAB(S_B) = \QSTAB(S_B)$. From the definition of $\HSTAB(H)$ and Prop. \ref{prop:sandhstab}, it follows that
    \begin{equation}
        \STAB(H) = \HSTAB(H),
    \end{equation}
    thus $\alpha(H,w) =  \hat{\alpha}(H,w)$. Then, the local and the broadcasting-local bound for $I(H,w)$ coincide.
\end{enumerate}
The latter case is particularly interesting: even if one, in principle, is weakening the constraints on the allowed probability distributions by considering only the constraint given by the \textit{exclusivity principle} on $\DAG_B$, one still retains an inequality whose bound is determined by the stronger condition of global local causality.

Moreover, if unitary weights are associated with $I(H,w)$, then $\hat{\alpha}(H)$ can be computed by noting that, due to the definition of $\HSTAB$, we have 
\begin{equation}
    \hat{\alpha}(H) = \max_{\omega_B \in \QSTAB(S_B)} \alpha(S_A, \omega_B),
    \label{eqn:weightahat}
\end{equation}
and in such case
\begin{equation}
    \hat{\alpha}(H) \leq \alpha(S_A),
    \label{eqn:hstab_unitary}
\end{equation}
which follows from the monotonicity of the independence number: $\alpha(S_A,w) > \alpha(S_A,w')$ if $w \succeq w'$. Thus, promising classes of inequalities (for a quantum violation) could be given by situations in which $\hat{\alpha}(H) \leq \alpha(S_A) \leq \theta(H)$.
Finally, the maximization necessary to compute the graph invariants, on positive weighted graphs can be carried out only on the extremal points of the related convex polytopes.
This fact, considering Prop.~\ref{prop:hstabvert}, leads us to the following consideration: while the set of extremal points of $\HSTAB(H)$ contains all the points associated with $\STAB(H)$ and with the local deterministic distributions, it contains only a subset of the extremal distributions associated with the exclusivity principle and $\QSTAB(H)$ \footnote{If it contained all, one would have that $\hat{\alpha} = \alpha^*$ leading to an inequality which cannot be used to test for the presence of a gap between the broadcast set and the quantum set.}.

Thus we can divide the candidate inequalities for a quantum violations into two categories:
i) the \emph{genuine} inequalities, where the broadcasting-local bound is strictly larger that the local one, which are associated with a weighted graph $(H,w)$, such that:
\begin{equation}
    \alpha(H,w) < \hat{\alpha}(H,w) < \theta(H,w),
    \label{eqn:genuine}
\end{equation}
and ii) the \emph{non-genuine} inequalities, when
\begin{equation}
    \alpha(H,w) = \hat{\alpha}(H,w) < \theta(H,w),
    \label{eqn:nongenuine}
\end{equation}
that is, the broadcasting-local and the local bound coincide.

To give a geometrical intuition behind the idea just presented, note that an inequality can always be thought of as a hyperplane ``cutting'' the relevant space of behaviors. 
As depicted in Fig. \ref{fig:genuine}, on one hand, there are situations in which $\hat{\alpha}$ corresponds to a facet connecting deterministic points - leading to an inequality for which the local bound and the broadcasting-local bound coincide. 
On the other, it is possible that the relevant facet of $\HSTAB$ connects extremal points fulfilling the EP, cutting the quantum correlation set given by $\mathrm{TH}$: in this instance, one would have a gap between the broadcasting-local set and the local set.
By finding an instance in which the latter situation arises, would show that in a hybrid scenario, a stronger and non-trivial notion of nonclassicality occurs: that is, it could not be traced back to the emergence of a local-quantum gap akin to a multipartite Bell scenario. 
Finally, we note that whenever $\theta(G,w) < \alpha^*(G,w)$ strictly, shows that tripartite generalized probabilistic theories (GPT)~\cite{hardy2001quantum,barrett2007information,chiribella2010probabilistic} generate a set of correlations violating quantum predictions. 
This implies that the inequality can be applied to distinguish quantum from GPT predictions.

In the next section we will present an instance of each class, while other relevant examples are reported in Appendix~\ref{apx:examples}.

\section{Examples of inequalities derived from hybrid graphs}
\label{sec:examples}
\subsection{Non-genuine inequality}
\begin{figure}[ht]
    \centering
    \includegraphics[width=1\columnwidth]{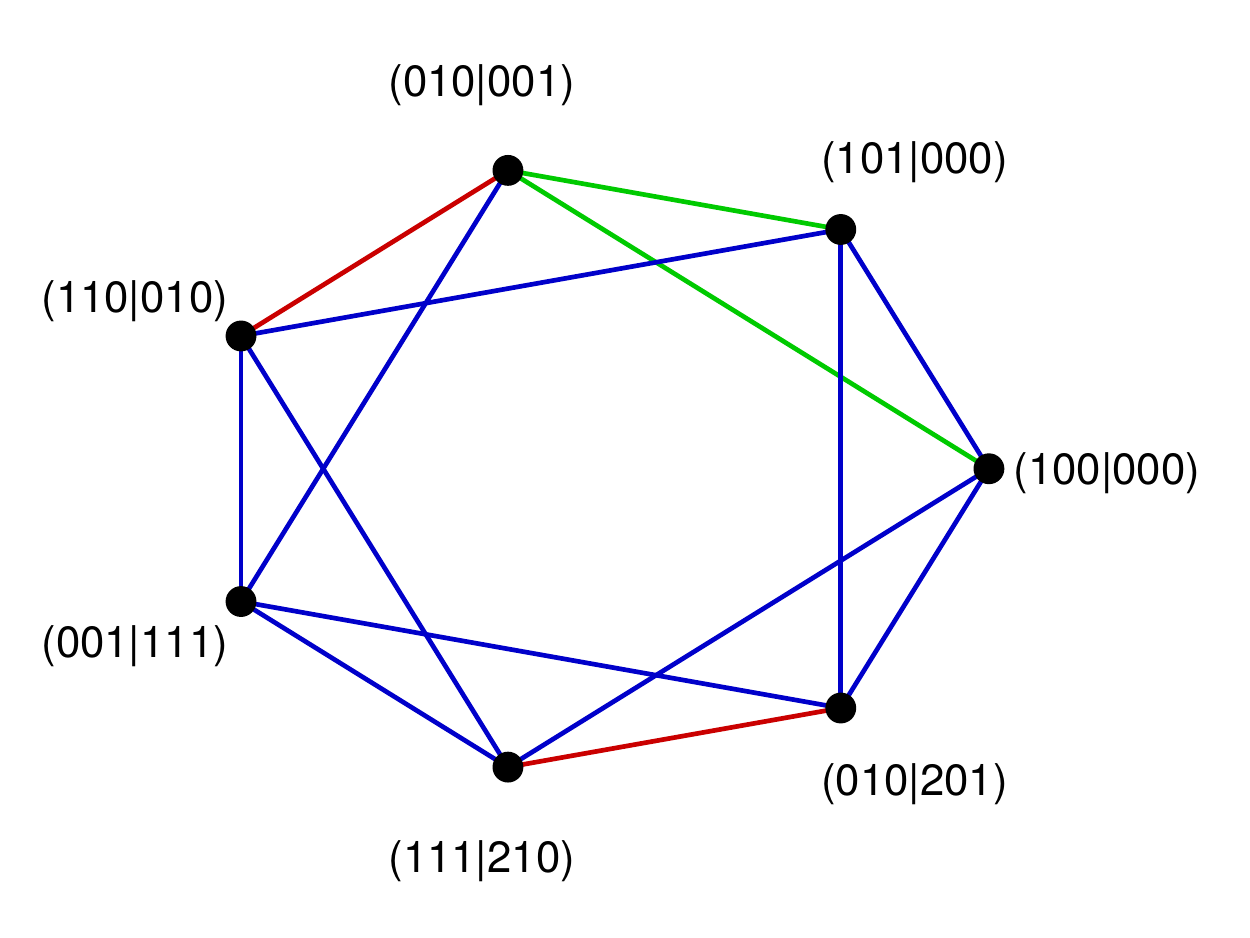}
    \caption{\textbf{Circulant graph inequality} associated with a circulant graph $H_1$ in the standard broadcasting scenario with $|X| =3$ and $|Y_1| = |Y_2| = 2$.
    This expression retains the same upper bound in the local and the broadcasting-local case $\alpha(H_1) = \hat \alpha(H_1) = 2$, making it an example of a \emph{non-genuine} inequality. At the same time this bound can be violated with a quantum resource up to $\beta_Q \approx 2.0697$.
    }
    \label{fig:circ7}
\end{figure}

A simple instance that falls in the first category can be constructed in the standard broadcasting scenario of Fig.~\ref{fig:broadcast}, with dichotomic measurements and $|X| =3$settings for $A$ and $|Y_1| = |Y_2| = 2$ settings for $B_1, B_2$.
The associated graph $H_1$, represented in Fig.~\ref{fig:circ7} is a so-called \emph{circulant graph} $\Ci_{1,2}(7)$, where $\Ci_{l,m}(n)$ correspond to a graph with $n$ vertex with ciclic edges $(i, i+l \mod n)$ and $(i, i+m \mod n)$.

By noticing that $\Ci_{1,2}(7) = \bar{C_7}$ is the complement to a $7$-cycle graph, from the SPGT~\ref{th:spgt}, we already expect the possibility of having a non trivial value for $\theta(\Ci_{1,2}(7))$. 
Indeed it can by using the fact that 
\begin{equation*}
    \theta(C_n) = \frac{n \cos(\pi/n)}{1 + \cos(\pi/n)}
\end{equation*}
valid for any cycle graph with odd $n > 3$, and the relation $\theta(G)\theta(\bar G) = n$~\cite{lovasz1979shannon, brimkov2000lovasz}, we conclude that 
\begin{equation}
    \theta(\Ci_{1,2}(7)) = 1 + \frac{1}{\cos (\pi/7)} \approx 2.109
    \label{eq:circ7_theta}
\end{equation}
Also it is straightforward to compute $\alpha(H_1) = \hat \alpha(H_1) = 2$.
This is also expected since $S_B$ is perfect, which means that we are in case 2 described in the previous section, where we have more generally that $\STAB(H_1) = \HSTAB(H_1)$.

This makes the inequality a \emph{non-genuine} one and at the same time a good candidate for a quantum violation.
It can be shown that there is a quantum strategy that violates the bound given by $\hat \alpha(H_1) = 2$ up to $\beta_Q \approx 2.0697$.
The full details of this realization are presented in appendix~\ref{apx:circ7_quantum}.
At the same time we can also upper bound the value of the quantum violation using the NPA hierarchy (computed at level $4$), which gives $\beta_Q \le 2.0697$, confirming that this represents the maximum quantum value.

\subsection{Genuine inequality}
\begin{figure}[ht]
    \centering
    \includegraphics[width=1\columnwidth]{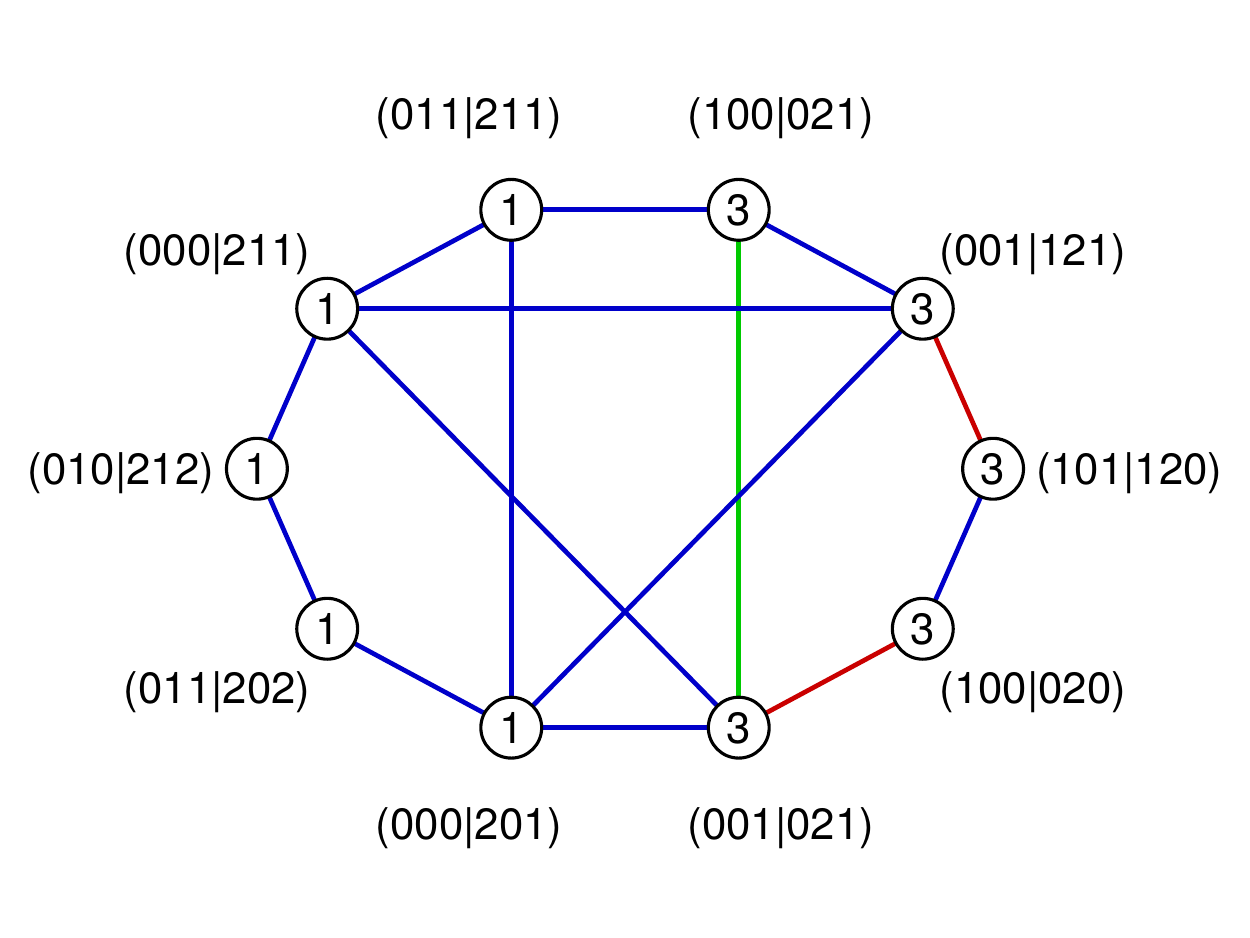}
    \caption{
    \textbf{Genuine inequality} graph $H_2$, with non-unitary weights $w_2$, present as an induced subgraph of the broadcasting scenario with $|X| = |Y_1| = |Y_2| = 3$.  
    For this set of events we have a strict inclusion $\STAB(H_2) \subset \HSTAB(H_2) \subset \TBODY(H_2)$, as shown by the values of the corresponding graph invariants.
    Moreover numerical test using the NPA hierarchy show that a quantum violation of the broadcasting-local bound is likely possible.
    }
    \label{fig:shell}
\end{figure}

We can also construct an example of an inequality of type ii) by considering the scenario with thre settings for each party, i.e. $|X| = |Y_1| = |Y_2| = 3$.
An instance of the corresponding graph $H_2$ is depicted in Fig.~\ref{fig:shell}, where in this case we are also using non-unitary weights $w_2$, as showed in the figure.

The graph $H_2$ contains four 5-cycles $C_5$ as induced subgraphs, therefore we expect it to be a good candidate for quantum violation from the SPG (Thm.~\ref{th:spgt}).
This can be verified by numerically computing the corresponding invariants:
\begin{equation}
\begin{split}
    &\alpha(H_2,w_2) = 8 \\
    &\hat \alpha(H_2, w_2) = 8.5 \\
    &\theta(H_2,w_2) \approx 8.9379 \\
    &\alpha^*(H_2,w_2) = 10
\end{split}
\end{equation}
In this case, all the inclusions are strict, i.e $\alpha(H_2,w_2) < \hat \alpha(H_2, w_2) < \theta(H_2,w_2) < \alpha^*(H_2,w_2)$, so this inequality is able, in principle, to distinguish between all the corresponding four sets.

Moreover, using the NPA hierachy~\cite{navascues2008convergent} (up to level $4$), we can check that the quantum bound is $\beta_Q \le 8.5352 < \theta(H_2,w_2)$, showing that this particular inequality could also be able to separate the quantum set from $\TBODY(H)$.

\section{Discussion}

In this work, we proposed an extension of the exclusivity graph method to analyze causal structures presenting hybrid constraints on part of their variables, which we refer to as hybrid scenarios. In this view, it is essential to devise methods capable of characterizing this scenarios and its hybrid constraints with the goal of giving a description of the properties of the associated correlation set. We provided a solution to this problem by outlining a graph-theoretical construction that is able to capture the subtleties given by the existence of a ``global'' constraint of local causality and only a weaker assumption given by the \emph{exclusivity principle} applied between the causal variables on only one side of the scenario.
Within this framework, we have shown how to map typical elements of graph theory, graph invariants and their associated sets, into the concepts of probabilistic behaviors and numerical bounds of given linear inequalities respectively. These are indeed the main elements that have to be identified when studying the emergence of nonclassicality in a device-independent setting, i.e. by only relying on a set of causal relationships encompassed by a causal scenario.

We have shown how by mapping a hybrid causal scenario into an exclusivity graph, one is able to identify similarities between probabilistic behaviors compatible with such scenario and its local counterpart, where one assumes a full set of classical (local) constraints.
Since such construction holds with respect to any subgraph - which can be associated to a linear probabilistic inequality - we derive criteria, relying only on the topology of the underlying graphs, that have to be satisfied in order to have the possibility of finding a discrepancy between the broadcast-local correlation set and the quantum one.

In this view, it is possible to obtain classes of inequalities for which the local and the broadcasting-local bound coincide; moreover, we have shown that it is possible to find instances in which a non-trivial notion of nonclassicality arises, i.e. it cannot be traced back directly to the emergence of a local-quantum gap akin to a multipartite Bell scenario.
In particular we presented an interesting instance of such \emph{genuine} inequality, also showing that these bounds can likely be violated by quantum mechanics.
Due to its generality, our method can be used in the treatment of general hybrid scenarios, featuring an arbitrary number of parties and/or a greater variety of causal relationships among them. In these situations, our approach could lead to the identification of other classes of hybrid causal structures in which non-classical behavior arises, striving away from the originally proposed tripartite broadcasting scenario~\cite{bowles2020single}. Given that these scenarios are of increasing interest for near-term applications, this extension can represent a tool for future realizations of these networks that will open the door for 
novel nonlocality demonstrations, both from a foundational \cite{villegasaguilar2023nonlocality} and an applied point of view.

\section*{Acknowledgements} 
This work was supported by the Serrapilheira Institute (Grant No. Serra-1708-15763), by the Simons Foundation (Grant Number 1023171, RC), the Brazilian National Council for Scientific and Technological Development (CNPq) via the National Institute for Science and Technology on Quantum Information (INCT-IQ) and Grants No. 307295/2020-6. This research was also supported by the Fetzer Franklin Fund of the John E.\ Fetzer Memorial Trust and by grant number FQXi-RFP-IPW-1905 from the Foundational Questions Institute and Fetzer Franklin Fund, a donor advised fund of Silicon Valley Community Foundation. We acknowledge support from the Templeton Foundation, through the ID 62312 grant, as part of the ‘The Quantum Information Structure of Spacetime - Second Phase’ Project (QISS 2). The opinions expressed in this project/publication are those of the author(s) and do not necessarily reflect the views of the John Templeton Foundation. We also acknowledge support from the ERC Advanced Grant QU-BOSS (QUantum advantage via nonlinear BOSon Sampling, grant agreement no. 884676).


\onecolumn
\appendix

\section{Mathematical definitions}
\label{apx:matdefi}

A subset $\mathds{R}^d_+$ is a convex polytope if:
\begin{defi}[Convex Polytope]
    A convex polytope is the convex hull of a finite set of points in $\mathds{R}^d$.
    \begin{equation}
        P := \chull\{x^1,x^2,...,x^n \in \mathds{R}^d\}.
    \end{equation}
\end{defi}
That is, a convex polytope $P \subset \mathds{R}^d_+$ is such that $x \in P$ if and only if it can be obtained as a convex combination of its extremal points. That is,
\begin{equation}
\begin{split}
    x \in P \Leftrightarrow x = \sum_i \alpha_i x^i, \\
    x^i \in \textrm{vert}(P) \; ; \; \sum_i \alpha_i = 1
    \label{eqn:decomp}      
\end{split}
\end{equation}
where with $\vertex(P)$ we denote the finite set of extremal points of $P$, defined as:
\begin{defi}[Extr. Point]
    $x \in P$ is an extremal point of a convex polytope $P$ if and only it only admits decompositions as in \eqref{eqn:decomp} where $\exists ! k$ such that $a_i = \delta_i^k$.
\end{defi}

Moreover the Hadamard product $\circ$ is defined as the following operation between vectors in $\mathds{R}^d_+$:
\begin{defi}[$\circ$-product]
    $\circ$ represents the element wise product between two vectors $x,y \in \mathds{R}^d_+$.
    \begin{eqnarray*}
    \begin{split}
     x = (x_1,x_2,...,x_d),  \\
    y = (y_1,y_2,...,y_d),\\
    z = x \odot y = (x_1y_1, x_2y_2, ..., x_dy_d) \in \mathds{R}^d_+,    
    \end{split}
    \end{eqnarray*}
\end{defi}

Furthermore, we note that defining the set:
\begin{equation}
    R := P \odot Q = \chull\{x \odot y \;|\; \forall (x,y) \;
    \textrm{s.t.} \; x \in \vertex(P); \; y \in \vertex(Q)\}.
\end{equation}
is not equivalent to taking the set of $x \circ y$ between all the vectors $x \in P$ and $y \in Q$: that is, $R$ might be larger than the latter. While $\forall x,y \in P,Q$ one has that
\begin{equation}
    x \odot y = (\sum_i \alpha_i x^i)\odot(\sum_j \beta_j y^j) = \sum_{i,j} \alpha_i \beta_j x^i \odot y^j \in R.
    \label{eqn:proof_conv}
\end{equation}
and since $\sum_{i,j} \alpha_i \beta_j = 1$, Eqn. \eqref{eqn:proof_conv} defines a convex combination of the finite set points used in \eqref{eqn:odot_poly} to define $R$. The converse is not necessarily true, since there exist convex combinations in the form
\begin{equation}
    z = \sum_i \gamma_i x^i \circ y^i \in R,
\end{equation}
such that
\begin{equation}
    z \neq x \circ y \; \textrm{with} \; x \in P; y \in Q.
\end{equation}

\section{Proof of Proposition \ref{prop:stabequiv}}
\label{apx:prop1}

\textbf{Proposition \ref{prop:stabequiv}}: {\em Given the hybrid exclusivity graph mapping described in the main text, the stable sets of the graphs $G_A$ and $G_B$ are such that:}
    \begin{equation}
        \STAB(G) = \STAB(G_A) \odot \STAB(G_B).
    \end{equation}
{\em Here with $G_A = (V_G, E_A)$ and $G_B = (V_G, E_B)$, we denote the exclusivity graphs as defined in Eqn. \eqref{eqn:egext_def}.}
\begin{proof}
Since both sides of the equality are defined as \textit{convex hulls}, we can restrict our analysis to the extremal points of the two sets.\\
($\rightarrow$) Both $E_A, E_B \subseteq E_G$ are proper subsets of the edge set of the full exclusivity graph $G$, it follows that if $\chi_S$ is a stable labeling for $G$, it is also a stable labeling of both $G_A$ and $G_B$. Thus we can write
\begin{equation*}
    \chi_S = \chi_A \circ \chi_B,
\end{equation*}
since one can take $\chi_S = \chi_A = \chi_B$ with $\chi_A \in \STAB(G_A)$, $\chi_B \in \STAB(G_B)$.\\
($\leftarrow$) By definition of $\STAB(G)$, it is enough to show that given $\chi_A \in \vertex(\STAB(G_A))$, $\chi_B \in \vertex(\STAB(G_B))$ then
\begin{equation*}
    \chi_G = \chi_A \circ \chi_B,
\end{equation*}
is a proper stable labeling of $G$. Consider each and every possible pair $(u,v)$ of distinct vertices of $G$, and denote as $\chi^{(u,v)}$ the corresponding components of a given stable labeling of $G$. If $(u,v) \notin E_G$, then every possible assignment given by $\chi^{(u,v)}_A \circ \chi^{(u,v)}_B$ gives a valid assignment for a $\chi^{(u,v)}_G \in \STAB(G)$. If instead $(u,v) \in E_A$, $(u,v) \notin E_B$, then with respect to the two nodes we have that
\begin{equation*}
\begin{split}
    \chi_A^{(u,v)} \in & \; \{(1,0);(0,1);(0,0)\}, \\
    \chi_B^{(u,v)} \in & \; \{(1,1);(1,0);(0,1);(0,0)\}, \\
    \chi_G^{(u,v)} \in & \; \{(1,0);(0,1);(0,0)\},
\end{split}
\end{equation*}
by definition of the Hadamard product $\circ$, which shows again that $\chi^{(u,v)}_G \in \STAB(G)$, independently of the assignments given by $\chi_B$. The other possible cases, $(u,v) \notin E_A$, $(u,v) \in E_B$ or $(u,v) \in E_A$, $(u,v) \in E_B$ can be treated similarly. Since this holds for every pair of vertices $\forall u,v \in V_G$, we conclude that $\chi_G \in \STAB(G)$.
\end{proof}

This proves that our hybrid construction correctly reproduces the local polytope as in Eqn.~\eqref{eqn:trilocal_distrib}, in the EG framework the set $\STAB(G)$, by considering separately two graphs $G_A$ and $G_B$ that embed in their edges two different kinds of exclusivity relations. Intuitively, this can be also shown as follows. Assuming that every variable $\{A,X;B_1,B_2;Y_1,Y_2\}$ takes values in $\{0,1\}$, then every extremal point of $\STAB(G_A)$ and $\STAB(G_B)$ associated with a maximal stable set represents - seen from the probabilistic perspective - deterministic strategies as:
\begin{equation*}
    D(A|X) = \mqty(D(0|0) \\ \vdots \\ D(1|1)) \in \vertex(\STAB(G_A)); \qquad
    D(B_1 B_2 | Y_1Y_2) = \mqty(D(00|00) \\ \vdots \\ D(11|11)) \in \vertex(\STAB(G_B)),
\end{equation*}
Then, by definition of the Hadamard product, one has:
\begin{equation*}
    D(A|X) \circ D(B_1B_2|Y_1Y_2) = \mqty(D(0|0)D(00|00) \\ \vdots \\ D(1|1)D(11|11)),
\end{equation*}
which is a valid deterministic strategy as in \eqref{eqn:det_local} and an extremal point of $\STAB(G)$. Since this reasoning can be applied to every pair of extremal points of $\STAB(G_A)$ and $\STAB(G_B)$, this shows that indeed:
\begin{equation}
        \STAB(G) = \STAB(G_A) \odot \STAB(G_B).
\end{equation}

\section{Proof of Proposition \ref{prop:hstabvert}}
\label{apx:prop2}

\textbf{Proposition \ref{prop:hstabvert}}
{\em The extremal points of $\HSTAB(G)$ form a subset of the extremal points of $\QSTAB(G)$, that is,}
\begin{equation*}
    \vertex(\HSTAB(G)) \subseteq \vertex(\QSTAB(G)).
\end{equation*}

\begin{proof}
Consider two extremal points of $\STAB(G_A)$, $\QSTAB(G_B)$ given by
\begin{eqnarray*}
\chi_A \in \vertex(\STAB(G_A)), \\
\omega_B \in \vertex(\QSTAB(G_B)),
\end{eqnarray*}
and the set of extremal points of $\QSTAB(G)$ given by
\begin{equation*}
    \{\omega_G^i\} \equiv \vertex(\QSTAB(G)),
\end{equation*}
By reductio ad absurdum, take $\chi_A$ and $\omega_B$ in such a way that
\[\chi_A \circ \omega_B \in \vertex(\HSTAB(G)).\]
and suppose that $\chi_A \circ \omega_B$ is \textbf{not} an extremal point of $\QSTAB(G)$, thus admitting a decomposition in the form
\begin{equation*}
    \chi_A \circ \omega_B = \sum_i \alpha_i \omega_G^i \quad \sum_i \alpha_i = 1; \quad \alpha_i > 0,
\end{equation*}
for more than one $i$. Since
\[\chi_A \circ \chi_A \circ \omega = \chi_A \circ \omega, \]
due to the fact that $\chi_A$ is a vector whose components are $\{0,1\}$, one can then write
\begin{equation*}
    \chi_A \circ \omega_B = \sum_i \alpha_i (\chi_A \circ \omega_G^i).
\end{equation*}
But since $E_B \subseteq E_G$, it is known that \cite{knuth1993sandwich}
\[\QSTAB(G) \subseteq \QSTAB(G_B),\]
because all the cliques of $G_B$ are still cliques of $G$, while the converse is not true; resulting in a larger set of constraints for $\QSTAB(G)$. Therefore, the extremal points of $\QSTAB(G)$ belong to $\QSTAB(G_B)$, that is,
\[\omega_G^i \in \QSTAB(G_B).\]
Recalling the definition of $\HSTAB(G)$, and the properties of the $\odot$ product amongst polytopes, this shows that, since
\[\chi_A \circ \omega_G^i \in \HSTAB(G),\]
then $\chi_A \circ \omega_B$ cannot be an extremal point of $\HSTAB(G)$, leading to a contradiction. This in turn implies that
\begin{gather*}
    \chi_A \circ \omega_B \in \vertex(\HSTAB(G)) \quad \Rightarrow \\
    \Rightarrow \quad \chi_A \circ \omega_B \in \vertex(\QSTAB(G)),
\end{gather*}
or alternatively that
\begin{equation*}
    \vertex(\HSTAB(G)) \subseteq \vertex(\QSTAB(G)).
\end{equation*}
\end{proof}

\section{Proof of Proposition \ref{prop:sandhstab}}
\label{apx:prop3}

\textbf{Proposition \ref{prop:sandhstab}}
{\em The set $\HSTAB(G)$ is \textit{sandwiched} between $\STAB(G)$ and $\QSTAB(G)$, that is,}
    \begin{equation*}
        \STAB(G) \subseteq \HSTAB(G) \subseteq \QSTAB(G).
    \end{equation*}
\begin{proof}
First, to prove that
\[\STAB(G) \subseteq \HSTAB(G),\]
we notice that since
\[\STAB(G) = \STAB(G_A) \odot \STAB(G_B),\]
using the fact that the sandwich theorem applied on $G_B$ gives
\[\STAB(G_B) \subseteq \QSTAB(G_B),\]
it follows directly from the definition of $\odot$ operation between polytopes that
\[\STAB(G) \subseteq \HSTAB(G). \]
On the other hand, due to theorem \ref{prop:hstabvert}, we know that
\[\vertex(\HSTAB(G)) \subseteq \vertex(\QSTAB(G)).\]
Since they are both convex polytopes defined as the convex hull of their extremal points, it follows that
\[\HSTAB(G) \subseteq \QSTAB(G).\]
\end{proof}

\section{Explicit quantum strategy for the $\bar C_7$ inequality}
\label{apx:circ7_quantum}
In section \ref{sec:examples} we presented an example of a non-genuine inequality for the scenario $|X|=3, |Y_1|=|Y_2|=2$.
Explicitly, in term of probabilities $p(a,b_1,b_2|x,y_1,y_2)$, the inequality is:
\begin{multline}
    S = p(1,0,0|0,0,0) +
    p(1,0,1|0,0,0) +
    p(0,1,0|0,0,1) +
    p(1,1,0|0,1,0) +\\+
    p(0,0,1|1,1,1) +
    p(1,1,1|2,1,0) +
    p(0,1,0|2,0,1) \le 2
    \label{eq:circ7_ineq}
\end{multline}

Here we describe two possible quantum realizations, using a tripartite state of three qubits and dichotomic measurement, that gives a violation of the inequality.

We consider first the case where the initial state is a GHZ state of the form $\ket{\psi} = (\ket{000} + \ket{111})/\sqrt{2}$.
Each party can then perform projective measurements on the states
\begin{align}
    \nonumber
    &\ket{A_i} = \cos(\alpha_{i})\ket{0} + e^{i\eta_{i}} \sin(\alpha_{i}) \ket{1}\\
    \nonumber
    &\ket{B_i} = \cos(\beta_{i})\ket{0} + e^{i\theta_{i}} \sin(\beta_{i}) \ket{1}\\
    &\ket{C_i} = \cos(\gamma_{i})\ket{0} + e^{i\zeta_{i}} \sin(\gamma_{i}) \ket{1}
    \label{eq:projectors_CC7_strategy}
\end{align}
where the index $i$ represent the corresponding setting.

Using these measurements we can show that a violation of~\ref{eq:circ7_ineq} up to $S = 2.042$ is achievable with the following parameters:
\begin{align*}
&\alpha_0 = 4/9 \pi,  &\eta_0 = 0\\
&\alpha_1 = 0,    &\eta_1 = 0\\
&\alpha_2 = 4/7 \pi,  &\eta_2 = 1/9\pi\\
&\beta_0 = 0,   &\theta_0 = 0\\
&\beta_1 = 2/9 \pi, &\theta_1 = -5/8\pi\\
&\gamma_0 = 2/9 \pi, &\zeta_0 = 5/9\pi\\
&\gamma_1 = 1/2 \pi, &\zeta_1 = 0
\end{align*}

This bound is still strictly lower than the upper bound computed directly with the NPA hierarchy method (up to level 4), corresponding to $S = 2.069$.
To reach this bound, we need a more general state of the form:
\begin{equation}
    \ket{\psi} = \sum_{ijk} \alpha_{ijk} e^{i\theta_{ijk}}\ket{ijk}
\end{equation}
with
\begin{align*}
   &\alpha_{000} = 0.4890 &\theta_{000} = \pi/2\\
   &\alpha_{100} = -0.1814 &\theta_{100} = 0 \\
   &\alpha_{010} = 0.5779  &\theta_{010} = 3/2 \pi \\
   &\alpha_{110} = -0.0687  &\theta_{110} = 3/2 \pi \\
   &\alpha_{001} = -0.0699  &\theta_{001} = 0 \\
   &\alpha_{101} = 0.5287  &\theta_{101} = \pi/7 \\
   &\alpha_{011} = 0  &\\
   &\alpha_{111} = -0.3240 &\theta_{111} = 0
\end{align*}

Using this state we can reach the value of $S = 2.069$ using projective measurements of the form \eqref{eq:projectors_CC7_strategy}, with the following parameters:
\begin{align*}
&\alpha_0 = 3/2 \pi,  &\eta_0 = 0\\
&\alpha_1 = 0,    &\eta_1 = 0 \\
&\alpha_2 = 1/3 \pi,  &\eta_2 = 3/2 \pi\\
&\beta_0 = 7/23 \pi,   &\theta_0 = 1\\
&\beta_1 = 3/2 \pi, &\theta_1 = 0\\
&\gamma_0 = 1/4 \pi, &\zeta_0 = 3/2\pi\\
&\gamma_1 = 1/2 \pi, &\zeta_1 = 0 \, .
\end{align*}

\section{Reproducing Bowles et al. inequality~\cite{bowles2020single}}
\label{sec:reprod}

\begin{figure*}[ht!]
    \centering
    \includegraphics{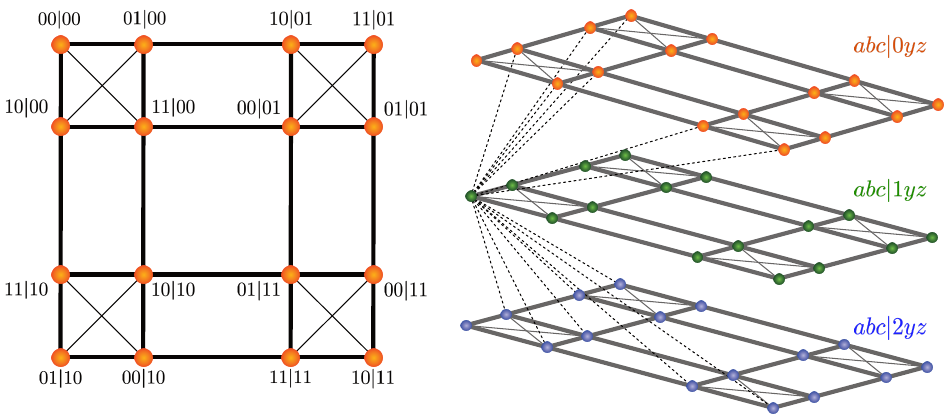}
    \caption{\textbf{Exclusivity Graph $S_B$ of Bowles inequality as in Eqn.\eqref{eqn:ineqbowl}:} on the left, the exclusivity graph associated with a standard $(2,2;2,2)$ Bell bipartite scenario; here bold lines represent cliques. On the right, Bowles' inequality \eqref{eqn:ineqbowl} in our framework is represented by three superimposed copies of such graph, each associated with different choices of Alice measurement setting $x$. Given an Alice's setting choice $x$, a node $\_ bc|xyz$ inherits an edge with each node, on a different 'plane', which is exclusive with an event $\_ b'c'|x'y'z'$ with respect to $bc|yz$. For clarity's sake, only one such connection is depicted with the dashed lines.}
    \label{fig:bowles}
\end{figure*}

In this section, we wish to show how, using the hybrid graph-theoretical method we introduced, it is possible to analyze an previous result, presented in~\cite{bowles2020single}, for the tripartite broadcasting scenario of Fig.~\ref{fig:broadcast}. 
In this scenario, it can be shown that the following inequality holds:
\begin{multline}
    \mathcal{I} = \expval{A_0B_0C_0} + \expval{A_0B_1C_1} + \expval{A_1B_1C_1} - \expval{A_1B_0C_0}+ \expval{A_0B_0C_1} + \expval{A_0B_1C_0} + \\ + \expval{A_1B_0C_1} - \expval{A_1B_1C_0} - 2\expval{A_2B_0} + 2\expval{A_2B_1} \overset{\text{B}}{\leq} 4
    \label{eqn:ineqbowl}
\end{multline}
where the expression is in terms of the correlators $\expval{A_xB_yC_z} = \sum_{a,b,c = 0,1} (-1)^{a+b+c} p(abc|xyz)$ and $\expval{A_xB_y} = \sum_{a,b = 0,1} (-1)^{a+b} p(ab|xy)$.
To employ our approach, one first has to transform \eqref{eqn:ineqbowl} into a linear combination of probabilities.
By doing this, one can show that \eqref{eqn:ineqbowl} can be mapped into an exclusivity graph $H$, with unitary weights, whose vertices are associated with the events listed in appendix~\ref{apx:bowles_ineq}.
 Moreover, we have that:
\begin{equation}
    \mathcal{I} \equiv I(H) \overset{\text{B}}{\leq} 8 \overset{\text{Q}}{\leq} 6 + 2\sqrt3 .
    \label{eqn:bowles_prob}
\end{equation}
To apply our construction, we analyzed separately the graphs $(S_A, S_B)$ containing edges due to $a|x$ and $bc|yz$, respectively. Interestingly, we notice that $S_A$ is constituted by three disjoint groups of $16$ nodes, each corresponding to a setting $x$ and equally split between $0|x$ and $1|x$. For our purposes, one can easily see that the extremal points of $\STAB(S_A)$ assign either
\begin{equation}
\begin{split}
    \omega_{0|x} = 1 \qquad \omega_{1|x} = 0 & \; \text{ or}\\
    \omega_{0|x} = 0 \qquad \omega_{1|x} = 1 &
\end{split}
\end{equation}
for each $x = 0,1,2$. 
Moreover, for each given $x$, all possible sub-events $bc|yz$ are present and the graph $S_B$ is essentially three superimposed exclusivity graphs of a standard bipartite $(2,2;2,2)$ Bell scenario, as shown pictorially in Fig.~\ref{fig:bowles}. 
Thus, due to the correspondence between the NS set and the $\QSTAB$ in a bipartite scenario~\cite{cabello2014graph}, the proper extremal points of $\QSTAB(S_B)$ will be given by the eight known symmetries of a PR-box \cite{popescu1994quantum,scarani2019bell}.
\begin{equation}
    \begin{split}
            \chi_{bc|00} &= \chi_{bc|01} = \chi_{bc|10} = \frac{1}{2}\delta_{b = c}, \\
            \chi_{bc|11} &= \frac{1}{2} \delta_{b \neq c} .
    \end{split}
    \label{eqn:PR_box}
\end{equation}
Using this fact, we immediately obtained all the relevant vertices of $\HSTAB(H)$, and one can indeed conclude that
\begin{equation}
    \alpha(H) = \hat{\alpha}(H) = 8.
\end{equation}
Also, we obtain:
\begin{equation}
    \theta(H) \approx 6 + 2 \sqrt 3,
\end{equation}
showing, that the quantum bound for the inequality~\eqref{eqn:ineqbowl} found by direct calculation in~\cite{bowles2020single} on a specific choice of state, transformation devices and measurements is in fact the maximum value achievable by quantum mechanics.

\subsection{Full list of events associated to Bowles et al. Inequality}
\label{apx:bowles_ineq}

The inequality presented in \cite{bowles2020single}:
\begin{multline}
    \mathcal{I} = \expval{A_0B_0C_0} + \expval{A_0B_1C_1} + \expval{A_1B_1C_1} - \expval{A_1B_0C_0}+ \expval{A_0B_0C_1} + \expval{A_0B_1C_0} + \\ + \expval{A_1B_0C_1} - \expval{A_1B_1C_0} - 2\expval{A_2B_0} + 2\expval{A_2B_1} \overset{\text{B}}{\leq} 4
\end{multline}
can be recast as a linear combination of $p(abc|xyz)$ with unitary coefficients and thus transformed in an exclusivity graph as:
\begin{equation}
\begin{split}
    \mathcal{I}(H,w) = p(000|000)+p(011|000)+p(101|000)+p(110|000)+p(000|011)+p(011|011)+p(101|011)+\\
    + p(110|011)+p(000|111)+p(011|111)+p(101|111)+p(110|111)+p(000|001)+p(011|001)+\\
    + p(101|001)+p(110|001)+p(000|010)+p(011|010)+p(101|010)+p(110|010)+p(000|101)+\\
    + p(011|101)+p(101|101)+p(110|101)+p(001|100)+p(010|100)+p(100|100)+p(111|100)+\\
    + p(001|110)+p(010|110)+p(100|110)+p(111|110)+p(000|210)+p(001|210)+p(110|210)+\\
    + p(111|210)+p(000|211)+p(001|211)+p(110|211)+p(111|211)+p(010|200)+p(011|200)+\\
    + p(100|200)+p(101|200)+p(010|201)+p(011|201)+p(100|201)+p(101|201) \overset{\text{B}}{\leq} 8
\end{split}
\end{equation}

\section{Examples of graph-based inequalities in hybrid scenarios}
\label{apx:examples}

\textbf{Mobius ladders in the tripartite broadcasting scenario}

\begin{figure}[ht]
    \centering
    \begin{tikzpicture}
        \foreach \i/\lbl in
        {1/(01|00),2/(11|02),3/(10|22),4/(01|22),5/(11|21),6/(10|11),7/(01|11),8/(10|10)}
            { \node [circle,fill=black] (\i) at ({-360/8 * (\i + 4.5)}:1.8) {};
              \node [] at ({-360/8 * (\i + 4.5)}:2.7) {\textcolor{black}{$\lbl$}};
            };
        \foreach \a/\b in {1/2,2/3,3/4,4/5,5/6,6/7,7/8,8/1}
            {
            \path (\a) edge [color=blue, ultra thick] (\b);
            };
    \end{tikzpicture}
    \begin{tikzpicture}
        \foreach \i/\lbl in
        {1/0|0,2/0|1,3/0|2,4/0|3,5/1|0,6/1|1,7/1|2,8/1|3}
            { \node [circle,fill=black] (\i) at ({-360/8 * (\i + 4.5)}:1.8) {};
              \node [] at ({-360/8 * (\i + 4.5)}:2.3) {\textcolor{red}{$\lbl$}};
            };
        \foreach \a/\b in {1/2,2/3,3/4,4/5,5/6,6/7,7/8,8/1}
            {
            \path (\a) edge [color=blue, ultra thick] (\b);
            };
        \foreach \a/\b in {1/5,2/6,3/7,4/8}
            {
            \path (\a) edge [color=red, ultra thick] (\b);
            };    
        \end{tikzpicture}
    
    \caption{\textbf{Example of an inequality represented by the M{\"o}bius ladder $M_8$:} consider a valid induced cycle for the $(3,3;2,2)_B$ bipartite scenario associated to $\DAG_B$, where each vertex represents an event in the form $b_1b_2|y_1y_2$, depicted on the left. Take this graph as $S_B$. Then, add simple edges related to $\{A,X\}$ in the form $0|x \sim 1|x$, as depicted on the right, to obtain a valid induced subgraph $H \simeq M_8$ in the broadcasting scenario. The associated inequality fulfills $I(H) \leq \alpha(H) = \hat{\alpha}(H) \leq \theta(H)$; with $\alpha(H) = 3$ and $\theta(H) = 3.4141$.}
    \label{fig:moebius}
\end{figure}
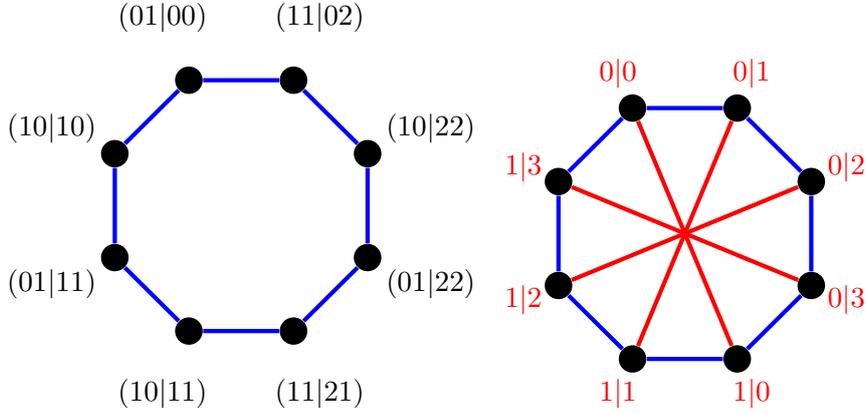

In the tripartite broadcasting scenario of Fig.~\ref{fig:broadcast}, following the prescriptions given in the main text we are able to find a class of inequalities, represented by subgraphs known in literature as \textit{M{\"o}bius Ladders}~\cite{guy1967mobius}. Notably, we can reverse the described procedure by considering at first a valid subgraph $S_B$ with respect to the exclusivity graph of $\DAG_B$, and then by 'attaching' to its nodes events $a|x$, thus building $S_A$ in the process.

We proceeded as follows: notice that in this case, we can interpret $\DAG_B$, as a bipartite scenario $(n,m;k,j)_B$, where
\begin{equation}
    n = |B_1| \quad m = |B_2| \quad k = |Y_1| \quad j = |Y_2|
\end{equation}
To give an analytical result, it is then natural to take $S_B$ as a generic induced cycle $C_s$ of the EG graph associated with an $(n,m;k,j)_B$ scenario of even cardinality and thus \textit{perfect}. We constructed and computationally analyzed such exclusivity graphs up to $n=m=9$, and for each combination of $(n,m)$, regardless of the values $(k,j)$, we were able to find induced cycles $C_s$ of cardinality:
\begin{equation}
    s \leq 3 \cdot \min(n,m)
\end{equation}

To construct a $2q$-vertices M{\"o}bius Ladder $M_{2q}$, notice that when considering only the nodes $\{A,X\}$, taking $|A| = 2$ and $|X| = q$, one is always able to consider a group of $q$ disjoint edges, given by:
\begin{equation*}
    \begin{split}
        0|0 \sim 1|0; \quad 0|1 \sim 1|1; \quad \ldots \quad 0|(q-1) \sim 1|(q-1) 
    \end{split}
\end{equation*}
Given this, with the following procedure one can construct an exclusivity graph $H$ as a M{\"o}bius ladder $M_{2q}$:
\begin{enumerate}
    \item Start with a cycle graph $C_{2q}$, defined on $S_B$, with even cardinality. This can be done if and only if $(n_B,m_B) > \lceil 2q/3 \rceil$.
    \item Associate to each diametrically opposed pair of vertices in $S_B$ the $q$ pairs of events $0|x$ and $1|x$.
\end{enumerate}
Notice that in this case, being $S_B$ topologically equivalent to a cycle graph $C_{2q}$ with an even number of vertices, it is a perfect graph. As stated in the main text, we then have that:
\begin{equation}
    \hat{\alpha}(H \simeq M_{2q}) = \alpha(H \simeq M_{2q})
\end{equation}
In Fig. \ref{fig:moebius}, we depict an inequality represented by M{\"o}bius ladder on eight vertices $M_8$. In this case, it is known that:
\begin{equation}
    \alpha(M_8) = 3 \qquad \theta(M_8) \approx 3.4141
\end{equation}
which shows a possibility for a quantum violation of the broadcasting-local bound.
Generally speaking, as done in \cite{bharti2021graph}, one can compute the values of $\alpha(M_{2q})$, verifying that:
\begin{equation}
    \alpha(M_{2q}) < \theta(M_{2q}) \quad \Leftrightarrow \quad 2q \mod{4} = 0 
\end{equation}
For a M{\"o}bius Ladder $M_{2q}$, with $q$ even, we then have a possible quantum-classical gap quantifiable as:
\begin{equation}
    \theta(M_{2q}) - \alpha(M_{2q}) = \frac{q}{2} \cos(\frac{\pi}{q}) - \frac{q}{2} + 1 \quad 
    \overset{q \rightarrow \infty}{\longrightarrow} \quad 1
\end{equation}

Employing the Python library \texttt{graph-tools}~\cite{peixoto_graph-tool_2014}, focusing on the $q=4$ case, we were able to find, up to a relabeling of the events $abc|xyz$, 16 unique inequalities. We proceeded to numerically optimize such inequalities over the quantum set up to the third level of the NPA hierarchy~\cite{navascues2008convergent}, finding that no actual quantum violation of the corresponding broadcasting-local bound is possible, since $\beta_{NPA} = \alpha$. This shows again that in the exclusivity graph framework, $\theta$ can be considered as a loose upper bound indicating a \textit{possible} quantum violation of an inequality, which can be falsified by a direct numerical optimization. 

\textbf{Circulant Graphs}
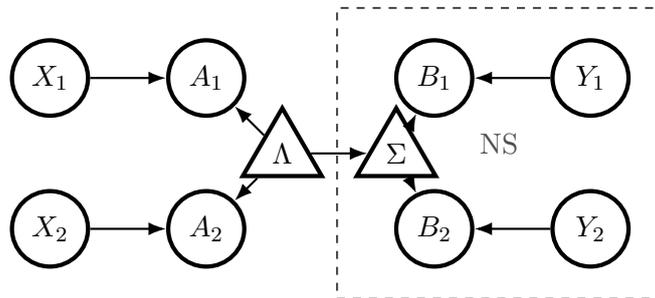
\begin{figure}[h]
    \centering
        \begin{tikzpicture}
         \node[var] (a1) at (-1,1) {$A_1$};
         \node[var] (a2) at (-1,-1) {$A_2$};
         \node[var] (x1) [left =of a1] {$X_1$};
         \node[var] (x2) [left =of a2] {$X_2$};
         \node[latent] (l) at (0,0) {$\Lambda$};
         \node[latent] (s) at (1.5,0) {$\Sigma$};
         \node[var] (b1) at (2,1) {$B_1$};
         \node[var] (b2) at (2,-1) {$B_2$};
         \node[var] (y1) [right =of b1] {$Y_1$};
         \node[var] (y2) [right =of b2] {$Y_2$};

         \path[dir] (x1) edge (a1) (x2) edge (a2) (y1) edge (b1) (y2) edge (b2);
         \path[dir] (l) edge (a1) (l) edge (a2) (l) edge (s);
         \path[dir] (s) edge (b1) (s) edge (b2);
         \node[thick, draw=black!70, color=black!70, dashed, fit=(s) (y1) (y2), inner sep=0.4cm] (ns)
         {$\NS$};
    \end{tikzpicture}
    \caption{\textbf{The 'double Bell' scenario:} Here a tripartite source distributes its system to two parties $A_1$ and $A_2$, which can be regarded as a bipartite Bell scenario $(n,m;k,j)_A$, while on the right side a device $\Sigma$ broadcasts the one share of the system to two parties $B_1$ and $B_2$, again forming a bipartite Bell scenario $(n,m;k,j)_B$.}
    \label{fig:double_bell}
\end{figure}

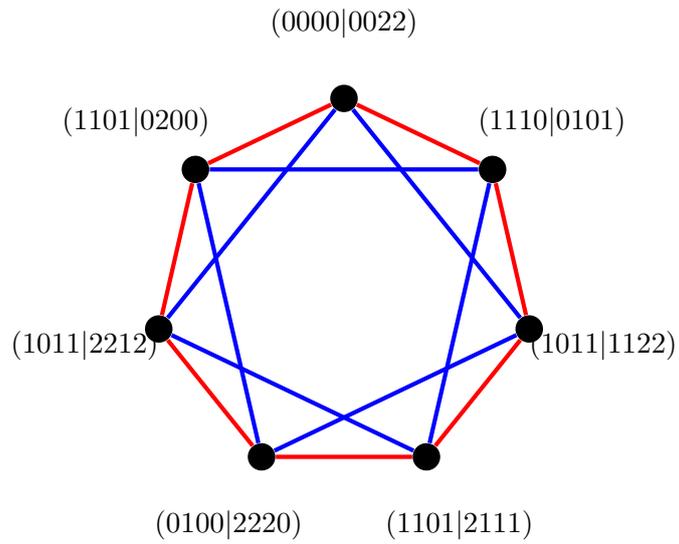
\begin{figure}[ht]
    \centering
    \begin{tikzpicture}
        \foreach \i/\lbl in
        {1/(0000|0022),2/(1110|0101),3/(1011|1122),4/(1101|2111),5/(0100|2220),6/(1011|2212),7/(1101|0200)}
            { \node [circle,fill=black] (\i) at ({-360/7 * (\i + 4.25)}:2.5) {};
            \node [] at ({-360/7 * (\i + 4.25)}:3.5) {$\lbl$};
            };
        \foreach \a/\b in {1/2,2/3,3/4,4/5,5/6,6/7,7/1}
            {
            \path (\a) edge [color=red, ultra thick] (\b);
            };
        \foreach \a/\b in {1/3,3/5,5/7,7/2,2/4,4/6,6/1}
            {
            \path (\a) edge [color=blue, ultra thick] (\b);
            };
    \end{tikzpicture}
    \caption{We depict a possible construction of an inequality represented by as circulant graph $C_7(1,2)$. Here the edges in red belong to $S_A$, while the edges in blue belong to $S_B$, both 7 nodes cycle graphs. With respect to this example of circulant graph, $S_A$ is depicted as an `outer' cycle, while $S_B$ as an `inner' cycle. Near each vertex we report the corresponding event $a_1a_2b_1 b_2|x_1x_2 y_1 y_2$ related to the causal structure depicted in Fig.~\ref{fig:double_bell}.}
    \label{fig:circ_construction}
\end{figure}

The causal scenario under consideration is depicted in Fig.~\ref{fig:double_bell}. It is obtained by considering as $\DAG_A$ and $\DAG_B$ two bipartite Bell scenarios. The global exclusivity graph can be constructed by considering exclusivity relations inferred on events in the form $a_1a_2b_1b_2|x_1x_2y_1y_2$ and applying the usual rules, previously described. In this case, the relevant broadcasting-local behaviour can be written as:
\begin{equation}
    p(a_1a_2b_1 b_2|x_1x_2 y_1 y_2) = \sum_\lambda p(\lambda) p(a_1|x_1,\lambda) p(a_2|x_2,\lambda) p_{B}(b_1 b_2|y_1 y_2,\sigma_\lambda)
\end{equation}
where again the only requirement on $p_{B}(b_1 b_2|y_1 y_2,\sigma_\lambda)$ is to be a no-signaling distribution as defined in Eqn.~\eqref{eqn:NS1}.
Within this structure, it is possible to devise a class of inequalities represented by so-called \textit{circulant} graphs, whose description is analytically possible due to their regularity.
\begin{defi}
A graph $G$ defined on the vertex set $\{v_i\}; \; i = \{1,2,...,n\}$ is a \textit{circulant} graph $C_n(1,k)$ if its edge set contains only edges in the form:
\begin{equation*}
\begin{split}
    (v_i,v_{i+1}) \in E_G \\
    (v_i,v_{i+k}) \in E_G
\end{split}
\end{equation*}
where $i + 1 \equiv (i+1) \mod{n}$ and $i + k \equiv (i+k) \mod{n}$
\end{defi}
Such instances arise by selecting $S_A$ and $S_B$ as cycle graphs constructed on events in the form$a_1a_2|x_1x_2$ and $b_1b_2|y_1y_2$ respectively, and 'intertwining' them, as in Fig.\ref{fig:circ_construction}.

Due to the regularity of circulant graphs, an analytical description of $\HSTAB$ is at reach. It is possible to show that either the cycle graph $S_B$ is even, and so \textit{perfect}, or it is odd, in which case the only proper fractional vertex of $\QSTAB(S_B)$ is given by $(1/2,\ldots,1/2)_q$, given the fact that an odd circulant graph is a \textit{minimally imperfect} graph~\cite{hoang1996some}. Through this, we are able to conclude, using Eqn.~\eqref{eqn:weightahat} and the known form of $\theta(C_q(1,2))$~\cite{brimkov2000lovasz}, that in this case indeed:
\begin{equation}
    \alpha(C_q(1,2)) = \hat{\alpha}(C_q(1,2)) = \floor{\frac{q}{3}} < \theta(C_q(1,2))
\end{equation}
which holds iff
\begin{equation}
     q\mod 3 \neq 0; \quad q \geq 7
\end{equation}
Even if also here $\alpha < \theta$, we found again that when such inequalities - one of which is reported in Fig.\ref{fig:circ_construction} - are numerically optimized up to the second level of the NPA hierarchy \cite{navascues2008convergent} 
one fails again to detect the presence of a nonclassical behaviour, since $\beta^2_{NPA} = \hat{\alpha}(C_q(1,2))$. Nonetheless, the strict inequality
\begin{equation}
    \hat{\alpha}(C_q(1,2)) < \alpha^*(C_q(1,2))
\end{equation}
holds. That is, even if no quantum-classical gap can arise, such inequalities could be used as post-quantum witnesses when confronted with resources constrained only by GPT theories.

\end{document}